\documentclass[twocolumn]{aastex631}
\usepackage{amsmath}

\begin{document}

\title{The Chandra Strong Lens Sample: Measuring the Dynamical States and Relaxation Fraction of a Sample of 28 Strong Lensing Selected Galaxy Clusters}

\author[0009-0004-7337-7674]{Raven Gassis}
\affiliation{Department of Physics, University of Cincinnati, Cincinnati, OH 45221, USA}

\author[0000-0003-1074-4807]{Matthew B. Bayliss}
\affiliation{Department of Physics, University of Cincinnati, Cincinnati, OH 45221, USA}

\author[0000-0001-5226-8349]{Michael McDonald}
\affiliation{Department of Physics, Massachusetts Institute of Technology, Cambridge, MA 02139, USA}
\affiliation{Kavli Institute for Astrophysics and Space Research, Massachusetts Institute of Technology, 77 Massachusetts Avenue, Cambridge, MA 02139, USA}

\author[0000-0002-7559-0864]{Keren Sharon}
\affiliation{Department of Astronomy, University of Michigan, 1085 South University Avenue, Ann Arbor, MI 48109, USA}

\author[0000-0003-3266-2001]{Guillaume Mahler}
\affiliation{STAR Institute, Quartier Agora -- All\'ee du six Ao\^ut 19c, B-4000 Li\`ege, Belgium}

\author[0000-0003-1370-5010]{Michael D. Gladders}
\affiliation{Department of Astronomy and Astrophysics, University of Chicago, 5640 South Ellis Avenue, Chicago, IL 60637, USA}
\affiliation{Kavli Institute for Cosmological Physics, University of Chicago, 5640 South Ellis Avenue, Chicago, IL 60637, USA}

\author[0000-0003-2200-5606]{H\aa kon Dahle}
\affiliation{Institute of Theoretical Astrophysics, University of Oslo, P.O. Box 1029, Blindern, NO-0315 Oslo, Norway}

\author[0000-0001-5097-6755]{Michael K. Florian}
\affiliation{Steward Observatory, University of Arizona, 933 North Cherry Ave., Tucson, AZ 85721, USA}

\author[0000-0002-7627-6551]{Jane R. Rigby}
\affiliation{Observational Cosmology Lab, Code 665, NASA Goddard Space Flight Center, Greenbelt, MD 20771, USA}

\author[0009-0007-6157-7398]{Lauren A. Elicker}
\affiliation{Department of Physics and Astronomy and PITT PACC, University of Pittsburgh, Pittsburgh, PA 15260, USA}

\author[0000-0002-2862-307X]{M. Riley Owens}
\affiliation{Department of Physics, University of Cincinnati, Cincinnati, OH 45221, USA}

\author[0009-0003-3123-4897]{Prasanna Adhikari}
\affiliation{Department of Physics, University of Cincinnati, Cincinnati, OH 45221, USA}

\author[0000-0002-3475-7648]{Gourav Khullar}
\affiliation{Department of Astronomy, University of Washington, Seattle, WA 98195, USA}

\begin{abstract}

We present the results of our dynamical state proxy measurements performed on 28 strong lensing galaxy clusters from the Sloan Giant Arcs Survey (SGAS). Using \emph{Chandra} ACIS-I/S X-ray data supplemented with \emph{HST} WFC3 imaging, we measure four morphological parameters: the concentration parameter (\emph{c}), asymmetry parameter (\emph{A}), centroid shift (\emph{log(w)}), and the X-ray–BCG centroid separation (\emph{D} [kpc]). Our goals are to (A) provide a robust classification of the dynamical state of the clusters in this strong lensing selected sample to enable studies that test various problems in cluster astrophysics and observational cosmology; (B) identify correlations, biases, or disagreements between different measurement proxies and cluster properties; and (C) measure the relaxation fraction (the fraction of clusters classified as relaxed based on X-ray morphology) and compare it to relaxation fractions from cluster samples selected using other methods.

We combine the four morphological parameters into a single metric, the combined parameter \emph{M}, which effectively separates the cluster sample into four dynamical state categories: relaxed; moderately relaxed; moderately disturbed; and disturbed. We find no significant trend in a cluster's dynamical state with its size, and only a weak, statistically limited dependence on mass and redshift. Based on our classification system, we find that $43\%^{+9}_{-9}$ of the clusters are relaxed, which is consistent with relaxation fractions measured for other cluster samples selected on mass-observables. This implies a strong lensing selected sample of clusters is on average dynamically similar to clusters selected via different methods.

\end{abstract}


\section{Introduction} \label{sec:intro}
\subsection{Impact of Dynamical State} \label{sec:dyn-st}
In our current understanding of cluster formation, galaxy clusters form through a process of hierarchical mergers in which smaller groups/clusters merge over time, eventually forming one large gravitationally bound object \citep{beers1983}. These objects are the most massive virialized systems, ranging from about $10^{13} M_\odot$ at the low-mass group scale up to about $10^{15} M_\odot$ at the high-mass cluster scale.

Due to the hierarchical accretion/merging mechanisms that govern cluster formation, a cluster undergoes various stages of dynamical activity during its formation \citep{Kravtsov2012}. Generally, we use a bimodal classification system to define a cluster's dynamical state, placing a given cluster into one of two presumably distinct populations: relaxed clusters or disturbed clusters. Some work has argued that a bimodal classification system is accurate \citep[e.g.,][]{cavagnolo2009,sanderson2009a,hudson2010}. Other work has found that a continuous distribution defining cluster dynamical state is more likely \citep[e.g.,][]{santos2010,pratt2010,ghirardini2017,yuan2020,campitiello2022,ghirardini2022}. If the latter is true, it makes separating clusters into two categories based on dynamical state more difficult. In the literature, most cuts on dynamical state are either arbitrary (as discussed in \citealt{cao2021}) or optimized using a specific sample which may or may not be representative of the true underlying cluster population.

Even if the underlying cluster population's dynamical state is characterized by a continuous distribution, cuts on dynamical state are necessary to conduct accurate scientific studies that require either relaxed or disturbed objects.

A well-defined sample of relaxed clusters is important to derive accurate masses for scaling relations \citep{pratt2009,lovisari2020}. The importance is more pronounced for certain mass proxies (e.g., X-ray hydrostatic methods) than others (e.g., lensing-based measurements), depending on their sensitivity to dynamical activity. The assumption of hydrostatic equilibrium only holds for relaxed systems; therefore, the resultant mass estimates from this assumption are inaccurate for dynamically active clusters. Thus, our understanding of cluster mass evolution and relation to other parameters directly depends on appropriate identification of these dynamically relaxed clusters \citep{johnston2007,george2012,czakon2015}.

On the other hand, a well-defined sample of disturbed clusters undergoing major mergers allows us to study the self-interaction of dark matter \citep{markevitch2004,harvey2015,monteiro-oliveira2018}, estimate non-thermal pressure support in clusters \citep{haiman2001,borgani2008,mantz2010}, investigate particle acceleration mechanisms resulting in non-thermal radio emission \citep{cassano2010,mann2012}, and other astrophysical phenomena that require the large-scale interaction only dynamically active clusters are able to create.

\subsection{Relaxation Measurements}\label{sec:rel-mes}

Observational data cannot directly measure the exact amount of dynamical activity in a galaxy cluster. However, observational proxies of different cluster components can provide insight into a cluster’s underlying dynamical activity.

A very widely studied cluster component to identify cluster dynamical state is the Intracluster Medium (ICM). Though all cluster components are sensitive to merger activity, the signatures of mergers persist on longer timescales in the ICM in the form of turbulence, shocks, cold fronts, clumps, and other inhomogeneities \citep{vikhlinin2001,markevitch2002,markevitch2005,markevitch2007,russell2010,nelson2014,rasia2014,eckert2015,biffi2016,planelles2017,ghirardini2018,ansarifard2020}. Many studies focus on X-ray maps of the ICM to measure morphological parameters that serve as proxy measurements for cluster dynamical state \citep[e.g.,][]{buote1995,lotz2004,santos2008,nurgaliev2013,weismann2013,wen2013,mantz2015,parekh2015,lovisari2017,mcdonald2017,nurgaliev2017,rossetti2017,bartalucci2019,yuan2020,ghirardini2022}. A plethora of different morphological indicators have been introduced and tested specifically using X-ray maps of the ICM that generally fall into two categories: those that focus on identifying a cool core, typically characterized by relatively high central gas density, and those that assess substructure throughout the cluster, from the core (sometimes excluded) to the outskirts (usually $\leq R_{500}$ where $R_{500}$ is the radius at which the mean density of a cluster is 500 times the critical density of the universe at the cluster's redshift). Additional studies have investigated the usefulness of Compton-y parameter maps (y-maps), which trace the thermal Sunyaev–Zel’dovich (SZ) effect, to study the morphology of the ICM in combination or in place of the analysis of X-ray maps \citep[e.g.,][]{cialone2018,capalbo2021,deluca2021}.

Studies have shown that optical indicators correlate with ICM morphological parameters and can also accurately distinguish cluster dynamical state \citep[e.g.,][]{sanderson2009b,mann2012,mahdavi2013,cui2016,rossetti2016,lopes2018,roberts2018,zenteno2020,seppi2023}. Typically, studies that incorporate optical measurements in their dynamical state analysis focus on how the Brightest Cluster Galaxy (BCG) behaves relative to the cluster at large, such as the offset between the BCG and cluster center \citep[e.g.,][]{lopes2018}, or to other member galaxies, such as the relative magnitudes of the BCG and $2^{nd}$ brightest cluster member \citep[e.g.,][]{seppi2023}.

\subsection{Strong Lensing Galaxy Clusters}\label{sec:slgc}

Recently, \citet{cerini2023} demonstrated that combining gravitational lensing and X-ray morphology can improve estimates of cluster dynamical state. Their analysis compared the power spectrum of the matter distribution derived from strong+weak lensing with that of the gas distribution characterized by ICM emission detected in Chandra X-ray data. Although limited in its sample size, this work illustrates the potential of incorporating gravitational lensing to probe cluster dynamics. However, the applicability of lensing-based dynamical state characterization of the general cluster population is limited due to data requirements.

In this study, we focus on clusters selected based on strong gravitational lensing. Using a strong lensing–selected sample allows us to assess how lensing-related selection biases affect the fraction of dynamically relaxed clusters in our sample. Simulations from \cite{Meneghetti2010b,Meneghetti2010a,Meneghetti2011} suggest that strong lensing clusters are uniquely biased objects. For instance, the lensing cross-section of a cluster is theorized to increase following recent merger activity \citep{torri2004,fedeli2006,hennawi2007,zitrin2013}. In fact, many observed strong lenses are dynamically complex \citep[see][]{lotz2017,jauzac2018}. However, lensing does not have a direct correlation to dynamical state because many cluster properties can correlate with increased lensing efficiency. \cite{hennawi2007} explored the role concentration plays in lensing efficiency. Concentration primarily affects the lensing efficiency of clusters at the lower mass end. More relaxed clusters can typically form denser cores, causing a relaxation bias for a subset of strong lenses. To explore these potential biases, we use X-ray morphological measurements commonly used on cluster samples to characterize the difference between a strong-lensing selected sample compared to samples with other selection functions.

\subsection{Paper Outline}
This paper aims to robustly define the dynamical state of 28 strong lensing galaxy clusters using a combination of standardized morphological indicators on the ICM and BCG; compare the parameters to each other and to cluster properties to identify correlations, biases, disagreements, or any other possible trends; and measure the fraction of relaxed clusters (relaxation fraction: $f_{\mathrm{relaxed}}$; see equation \ref{eq:f}) and compare our $f_{\mathrm{relaxed}}$ to those from cluster samples selected using other methods. In section~\ref{sec:dat-meas}, we describe our sample data and the dynamical state indicators used in this paper. In section~\ref{sec:results}, we present our results and discuss the correlation between the different measurement parameters. In section~\ref{sec:disc}, we discuss in more detail the correlations our measurements have with other cluster properties and how our relaxation fraction compares to those derived for other cluster samples.  We summarize our major findings in section~\ref{sec:sum}. In this paper, we assume a flat $\Lambda$CDM cosmology with $\Omega_\Lambda = 0.7$, $\Omega_m = 0.3$, and $H_0 = 70$ $\mathrm{km}$ $\mathrm{s}^{-1}$ $\mathrm{Mpc}^{-1}$.

\begin{figure*}
	\includegraphics[width=\textwidth]{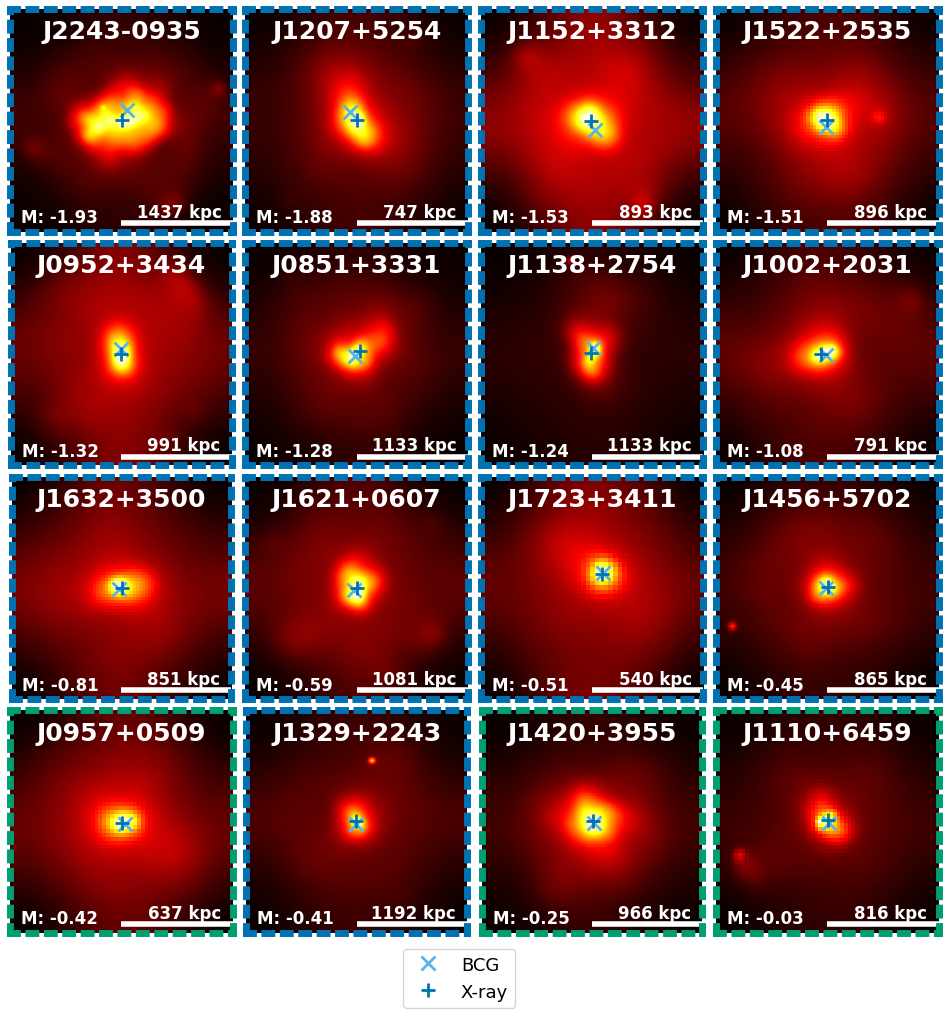}\centering
    \caption{Adaptively smoothed X-ray images of our disturbed cluster sample using the \texttt{CIAO} \texttt{csmooth} function. The objects with blue borders are those identified as ``disturbed" and the objects with green borders are those identified as ``moderately disturbed." The objects are sequenced according to their combined dynamical relaxation M parameter value (see section \ref{sec:M}) increasing from left to right and top to bottom.}
    \label{fig:dist_vis}
\end{figure*}

\begin{figure*}
	\includegraphics[width=\textwidth]{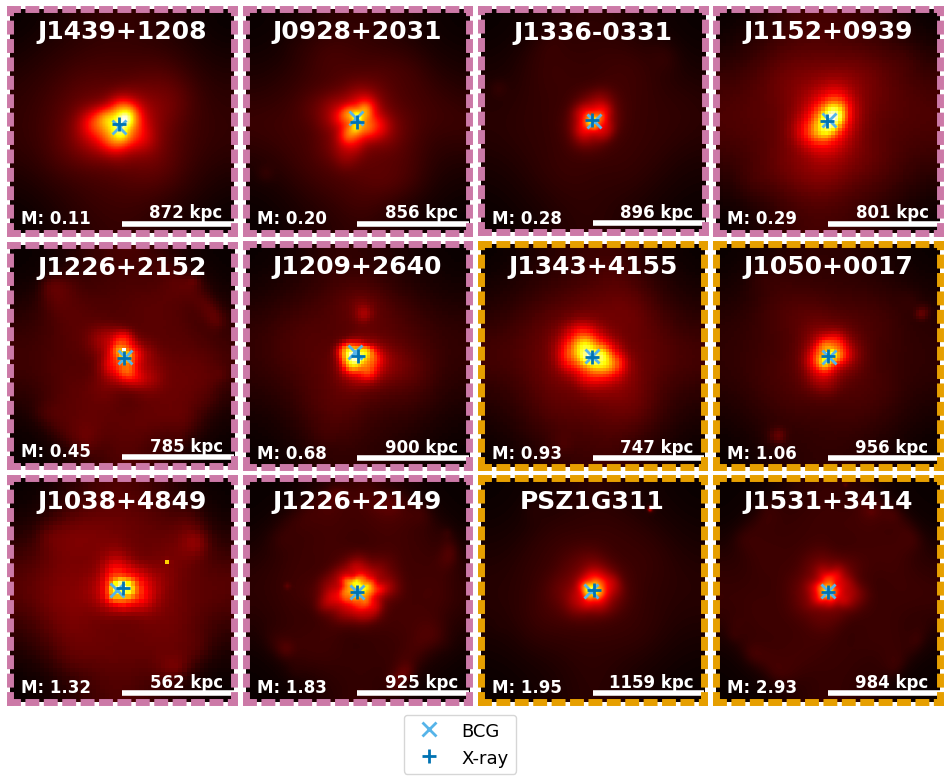}\centering
    \caption{Adaptively smoothed X-ray images of our relaxed cluster sample using the \texttt{CIAO} \texttt{csmooth} function. The objects with orange borders are those identified as ``relaxed" and the objects with magenta borders are those identified as ``moderately relaxed." The objects are sequenced according to their combined dynamical relaxation M parameter value (see section \ref{sec:M}) increasing from left to right and top to bottom.}
    \label{fig:rel_vis}
\end{figure*}

\begin{deluxetable*}{ccccccccc}
\centerwidetable
\setlength{\tabcolsep}{8pt}
\def\arraystretch{1.2}
\tablecaption{This table includes the cluster observation details in addition to some general cluster properties. The right ascension $\alpha$ and declination $\delta$ are the target coordinates for the program observations. We additionally detail the exposure time ($t_{exp}$), counts within $R_{500}$ ($N_{counts}$), and observation IDs associated with the program. We also list $R_{500}$, $M_{500}$, and the cluster redshift, $z$. All observations were collected with \emph{Chandra} ACIS‑I/S.\label{tab:prop-obs}}
\tablehead{
\colhead{Name} & \colhead{$\alpha$} & \colhead{$\delta$} & \colhead{$z$} & \colhead{$R_{500}$ [kpc]} & \colhead{$M_{500}$ [$10^{14}\,M_\odot$]} & \colhead{$t_{\mathrm{exp}}$ [s]} & \colhead{OBSIDs} & \colhead{$N_{counts}$}
}
\startdata
J0851+3331 & 132.91708 & +33.52314 & 0.3690 & $1134^{+158}_{-107}$ & $5.4^{+2.6}_{-1.7}$ & 11427 & 04205 & 3786 \\
J0928+2031 & 142.01875 & +20.52922 & 0.1920 & $856^{+3}_{-11}$ & $2.15^{+0.02}_{-0.08}$ & 6634 & 11767 & 5116 \\
J0952+3434 & 148.16667 & +34.57972 & 0.3570 & $992^{+268}_{-129}$ & $2.8^{+3.3}_{-1.6}$ & 28092 & 20564,20981 & 3306 \\
J0957+0509 & 149.41250 & +05.15889 & 0.4480 & $638^{+11}_{-11}$ & $1.18^{+0.06}_{-0.06}$ & 27714 & 20565 & 1133 \\
J1002+2031 & 150.61250 & +20.51750 & 0.3190 & $791^{+7}_{-13}$ & $1.96^{+0.05}_{-0.10}$ & 20190 & 20566 & 2535 \\
J1038+4849 & 159.67971 & +48.82194 & 0.4308 & $563^{+14}_{-7}$ & $0.79^{+0.06}_{-0.03}$ & 70199 & 11756,12098 & 5508 \\
J1050+0017 & 162.66667 & +00.28500 & 0.5931 & $957^{+206}_{-181}$ & $5.7^{+3.1}_{-2.7}$ & 34413 & 20567,22081 & 3035 \\
J1110+6459 & 167.57500 & +64.99639 & 0.6560 & $816^{+454}_{-76}$ & $3.1^{+5.4}_{-0.9}$ & 38406 & 20568,22099 & 2209 \\
J1138+2754 & 174.53750 & +27.90861 & 0.4510 & $1134^{+12}_{-9}$ & $6.73^{+0.22}_{-0.16}$ & 8042 & 20569 & 2113 \\
J1152+0930 & 178.19583 & +09.50389 & 0.5170 & $801^{+70}_{-89}$ & $2.48^{+0.67}_{-0.85}$ & 32156 & 20570 & 1945 \\
J1152+3312 & 178.00000 & +33.22833 & 0.3612 & $893^{+362}_{-159}$ & $2.78^{+3.6}_{-1.6}$ & 30675 & 20579 & 2936 \\
J1207+5254 & 181.89958 & +52.91639 & 0.2750 & $748^{+11}_{-16}$ & $1.56^{+0.07}_{-0.10}$ & 7945 & 18220 & 1916 \\
J1209+2640 & 182.34870 & +26.67960 & 0.5606 & $901^{+144}_{-44}$ & $3.85^{+1.83}_{-0.56}$ & 15875 & 18291 & 2117 \\
J1226+2149 & 186.71458 & +21.85431 & 0.4350 & $926^{+40}_{-29}$ & $3.59^{+0.47}_{-0.34}$ & 130237 & 12878 & 18046 \\
J1226+2152 & 186.71458 & +21.85431 & 0.4350 & $785^{+30}_{-28}$ & $2.17^{+0.25}_{-0.23}$ & 130237 & 12878 & 11117 \\
J1329+2243 & 202.39167 & +22.71667 & 0.4427 & $1193^{+263}_{-190}$ & $5.24^{+5.13}_{-3.70}$ & 18141 & 20572 & 2770 \\
J1336$-$0331 & 204.00000 & –03.52472 & 0.1764 & $896^{+540}_{-91}$ & $2.53^{+4.40}_{-0.74}$ & 9160 & 20573 & 3566 \\
J1343+4155 & 205.88667 & +41.91761 & 0.4180 & $748^{+5}_{-19}$ & $1.84^{+0.04}_{-0.14}$ & 16802 & 18218 & 2587 \\
J1420+3955 & 215.16250 & +39.91806 & 0.6070 & $967^{+148}_{-232}$ & $9.49^{+2.30}_{-3.61}$ & 26713 & 20580 & 3051 \\
J1439+1208 & 219.79167 & +12.14028 & 0.4273 & $872^{+16}_{-19}$ & $2.96^{+0.16}_{-0.19}$ & 13203 & 20575 & 1888 \\
J1456+5702 & 224.00417 & +57.03889 & 0.4840 & $866^{+157}_{-89}$ & $3.09^{+1.69}_{-0.96}$ & 26518 & 20576 & 2093 \\
J1522+2535 & 230.72083 & +25.59167 & 0.6020 & $896^{+295}_{-116}$ & $3.95^{+3.92}_{-1.55}$ & 23475 & 20578 & 1560 \\
J1531+3414 & 232.79417 & +34.24028 & 0.3350 & $985^{+99}_{-71}$ & $3.52^{+1.16}_{-0.83}$ & 122188 & 18689 & 34155 \\
PSZ1G311   & 237.51833 & –78.18333 & 0.4430 & $1160^{+173}_{-66}$ & $6.69^{+3.18}_{-1.22}$ & 39528 & 20442 & 9674 \\
J1621+0607 & 245.38333 & +06.12222 & 0.3429 & $1081^{+159}_{-91}$ & $4.99^{+2.26}_{-1.29}$ & 19499 & 20581 & 3886 \\
J1632+3500 & 248.04167 & +35.00833 & 0.4660 & $851^{+173}_{-158}$ & $3.31^{+1.79}_{-1.64}$ & 29693 & 20582 & 2057 \\
J1723+3411 & 260.90042 & +34.19939 & 0.4423 & $541^{+13}_{-15}$ & $0.72^{+0.05}_{-0.06}$ & 18798 & 18224 & 1022 \\
J2243$-$0935 & 340.83458 & –09.59111 & 0.4470 & $1437^{+45}_{-44}$ & $13.5^{+1.3}_{-1.3}$ & 20503 & 03260 & 9659 \\
\enddata
\end{deluxetable*}

\section{Data and Measurements}
\label{sec:dat-meas}

\subsection{Cluster Sample}\label{sec:samp}

We focus on a sample of 28 strong lensing galaxy clusters (see Table~\ref{tab:prop-obs}) with Chandra X-ray data that enables us to make measurements of the ICM components of the cluster sample. All clusters with the exception of PSZ1G311 are part of the Sloan Giant Arcs Survey (SGAS) \citep{koester2010,bayliss2011,bayliss2014,gladders2013,sharon2020} cluster sample. The SGAS clusters were selected based on the presence of giant arcs in SDSS imaging data. PSZ1G311 is additionally included since it also has giant arcs present in shallow ground-based imaging. Its selection, characteristics, and available data are similar to those of the SGAS clusters even though it is outside the SDSS footprint. All clusters have corresponding WFC3 imaging data from HST programs GO-13003 (PI: Gladders), GO-15378 (PI: Bayliss), GO-15377 (PI: Bayliss), or GO-12368 (PI: Morris). Of the 28 clusters with Chandra, 27 have well constrained lens models that allow us to construct mass maps of the underlying dark matter halo distribution. As discussed in \citealt{sharon2020}, SDSS J1002+2031 does not have sufficient data to produce a well-constrained lens model; still, it remains in this study due to its original selection based on its strong lensing features and sufficient X-ray follow-up data. The lens models are publicly available and discussed in \cite{sharon2020,sharon2022,sharon2023}.

For each cluster, we estimate $M_{500}$ (and $R_{500}$ correspondingly) from the X-ray imaging data. We derive M$_{500}$ and R$_{500}$ using the T$_X$—M relation from \citealt{Vikhlinin2009}, as described in detail in \citealt{McDonald2013} and \citealt{Andersson2011}. For each cluster, we iteratively measure a core-excised temperature in an aperture from $0.15{\times}R_{500}$ to $R_{500}$, centered on the large-scale X-ray centroid. This core-excised temperature is converted into a mass (M$_{500}$) via the T$_X$—M relation, which is then converted into a revised R$_{500}$. We iterate on this procedure until the value of R$_{500}$ changes by less than 5$\%$.

In the majority of clusters in our sample, the ICM is spatially coincident with a single optical core system composed of the BCG and the surrounding low surface brightness intracluster light (ICL) identified in the WFC3 imaging data. We find three systems in which two optical cores were found to be extremely close together with respect to redshift and position on the sky, indicating that they are in some stage of a merging process. The optical core pairs and their separations are listed in Table \ref{tab:2pairs}.

SDSS J0928+2031 has two BCG+ICL systems, one in the North-West and one in the South-East. The presence of two distinct BCG+ICL systems is clear in the optical data. However, in the X-ray data, the Northwestern core dominates the overall distribution of the ICM immensely. In consequence, we only consider the dominant northwestern core to characterize the dynamical state of this cluster.

SDSS J2243$-$0935 also has two BCG+ICL like systems, one in the East and one in the West. The ICM envelops both cluster cores and extends such that the centroid of the core is in a superposition between the two optical cores. To analyze the cluster as a singular object, we compare the overall ICM distribution to the western optical core which is reasonably identified as the dominant core. 

Both SDSS J0928+2031 and SDSS J2243$-$0935 appear to be at different stages of a significant major merger with the formerly separate ICM's fully interacting, therefore, we compare the dominant optical cores to the single ICM distribution for all morphological measures for the respective clusters. As will be noted later, in many situations we exclude SDSS J2243$-$0935 in the discussion of our results. For completeness, we include data points associated to SDSS J2243$-$0935 where appropriate.  

For SDSS J1226+2149 and SDSS J1226+2152, the ICMs of these two systems have yet to undergo a merging process. As such, these two systems are treated as two separate clusters despite their relative proximity. Detailed analysis of the possible interaction of these two clusters is beyond the scope of this paper. However, we note that SDSS J1226+2149 and SDSS J1226+2152 might be in a premerger state.

\begin{deluxetable}{ccc}
\setlength{\tabcolsep}{.045\columnwidth}

\tablecaption{Here we list the physical separation in units of kpc for the optical pairs in our sample that are close in redshift and position on the sky.The uncertainty in BCG positions is much smaller than the separation, so we only report the value of the separation.\label{tab:2pairs}}

\tablehead{\colhead{Core 1} & \colhead{Core 2} &  \colhead{Separation}} 

\startdata
J0928+2031 NW & J0928+2031 SE & $\sim$148 kpc \\ 
J2243$-$0935 W & JJ2243$-$0935 E & $\sim$399 kpc \\
J1226+2152 & JJ1226+2149 & $\sim$865 kpc \\
\enddata
\end{deluxetable}

\subsection{Chandra Data Reduction}
\label{sec:chan}

We made all of our ICM measurements using \emph{Chandra} ACIS-I/S X-ray event data. In figures~\ref{fig:dist_vis} and~\ref{fig:rel_vis}, we display the adaptively smoothed images of our data for the purpose of visualization. For actual measurements, we opt for a more robust data reduction process that should not be sensitive to the smoothing choices that bias measurement results \citep{yuan2020}.
We processed the raw \emph{Chandra} X-ray event data using the publicly available \texttt{CIAO} tools for data reduction. As suggested by the \texttt{CIAO} analysis threads, we started our reduction by reprocessing the data using \texttt{chandra\_repro}. We then looked at each observation to identify point sources. To do this, we used \texttt{mkpsfmap} to determine the Point Spread Function (PSF) at each individual pixel in the event file then used \texttt{wavdetect} which identifies point sources by correlating ``Mexican Hat" wavelet functions to the processed X-ray event files. We also manually identified the chip gaps in our event files. The point sources and chip gaps were catalogued in \texttt{ds9} region files and filled in using \texttt{dmfilth} using the POISSON mode to draw values. After, we use \texttt{merge\_obs} using the broad energy band (0.5 keV to 7 keV with an effective energy of 2.3 keV) to create the final output images for our ICM morphology measurements. In addition, we combine the region files from multiple observations and transform it into a single mask image. As detailed in Table \ref{tab:prop-obs}, the exposure times vary greatly across our sample; however, all clusters have sufficient counts within $R_{500}$ ($N_{counts}>1000$) such that we are able to measure the morphological properties of interest confidently.

\begin{figure}
	\includegraphics[width=.9\columnwidth]{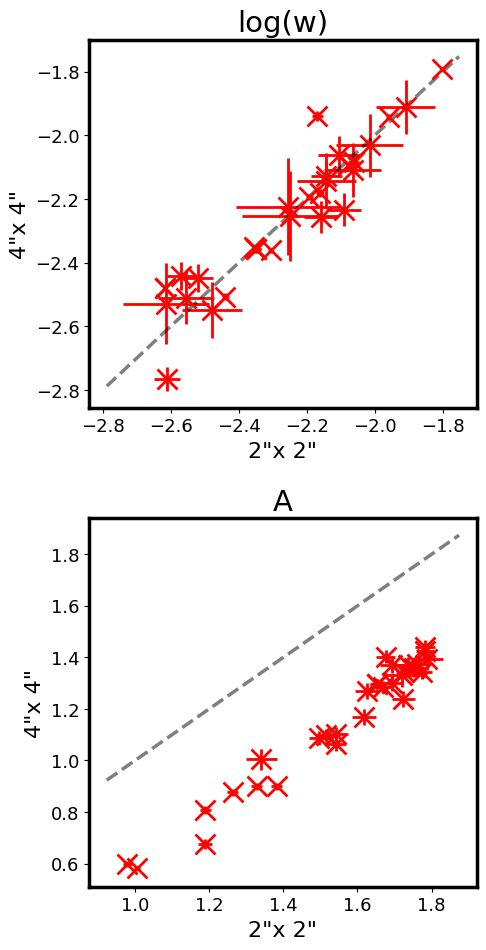}\centering
    \caption{A test of the effect of binning choice on the \emph{log(w)} (top panel) and \emph{A} (bottom panel) parameters. Each is calculated using unsmoothed images binned by either$2''\times2''$ (x-axis) or $4''\times4''$ (y-axis). For reference we plot the y=x line. We see that \emph{log(w)} is insensitive to binning whereas \emph{A} changes systematically based on choice of bin.}
    \label{fig:bin_comp}
\end{figure}

\subsection{ICM Morphological Measures}\label{sec:icm-meas}

\cite{lovisari2017} showed that it is best practice to combine parameters sensitive to substructure with parameters sensitive to the core properties. To this end, we analyze two parameters sensitive to substructure, the asymmetry parameter (\emph{A}) and centroid shift parameter (\emph{log(w)}), and two parameters sensitive to the core (the concentration parameter (\emph{c}) and X-ray-bcg centroid separation (\emph{D} [kpc])). We are able to measure these parameters directly in the X-ray images produced from the Chandra data reduction.

\subsubsection{Asymmetry and Centroid Shift}\label{sec:a-w}

As mentioned in section~\ref{sec:rel-mes}, signatures from merger activity appear as inhomogeneities in the ICM. These inhomogeneities can manifest themselves in detectable substructures that form in different regions of the ICM. Therefore, a good test of recent merger activity is to probe the extended space of the ICM to find these substructures. We do this using the \emph{A} and \emph{log(w)} parameters.

The asymmetry parameter (\emph{A}) has been used in previous studies to effectively differentiate between relaxed and disturbed clusters \citep[e.g.,][]{schade1995,okabe2010,zhang2010,rasia2013}. We measure \emph{A} by rotating the initial X-ray distribution $I$ by $180^\circ$ and subtracting the rotated distribution $R$ from the original distribution and taking the sum of the difference normalized by the sum of the initial distribution as illustrated by equation~\ref{eq:A}.

\begin{equation}
A=\frac{\sum{(|I-R|)}}{\sum{I}}
\label{eq:A}
\end{equation}

We computed \emph{A} using an aperture of $R_{500}$. When computing \emph{A}, we mask out the CCD chip gaps and point source pixel points instead of using the Poisson filled images detailed in \ref{sec:chan} to minimize spurious asymmetries. The greater the value we measure for \emph{A}, the more disturbed the cluster is. 

The centroid shift parameter (\emph{log(w)}) also detects substructure using an iterative approach. The definition of \emph{w} is given by equation~\ref{eq:w} where $\Delta_i$ is the separation of centroids between the centroid computed within $R_{500}$ and the centroid computed within the $i^{th}$ aperture. Here, $N$ is the total number of apertures. We take the logarithm of equation~\ref{eq:w} for our final determination of the centroid shift.

\begin{equation}
w=\frac{1}{R_{500}}\times\sqrt{\frac{\sum{(\Delta_i-\langle\Delta\rangle)^2}}{N-1}}
\label{eq:w}
\end{equation}

We step down $\Delta_i$ from $R_{500}$ to $0.15{\times}R_{500}$ in increments of $0.05{\times}{R_{500}}$. As we change the aperture used to estimate the centroid, we change the amount of substructure potentially influencing the determination of our centroid. As such, the greater the value of \emph{log(w)} the more disturbed a cluster is likely to be.

We tested how binning affects the value measured by \emph{A} and \emph{log(w)}. When comparing the morphological parameters measured on $2''\times2''$ and $4''\times4''$ binned \emph{Chandra} images, we find that the asymmetry parameter (\emph{A}) differs on average by 26.4\%, with a maximum difference of 43.2\%, and the centroid shift parameter (\emph{log(w)}) shows a much smaller average difference of 2.4\%, with a maximum of 10.6\%. In Figure~\ref{fig:bin_comp}, we see that \emph{log(w)} stays around the one to one line with minimal scatter and that \emph{A} is translated down from the one to one line by a constant. We opt for the $4''\times4''$ binned images because the broader binning suppresses Poisson noise more effectively than the $2''\times2''$ images, enabling more reliable identification of substructure. We do not bin beyond $4''$ since further binning could reduce our spatial resolution and wash out small scale features.


\begin{deluxetable*}{ccccc}
\setlength{\tabcolsep}{16pt}
\def\arraystretch{1.2}

\tablecaption{The BCG and X-ray centroids. $\alpha$ and $\delta$ are the right ascension and declination reported in degrees (J2000).\label{tab:ra-dec}}

\tablehead{\colhead{Name}  & \colhead{$\alpha_{BCG}$} & \colhead{$\delta_{BCG}$} &  \colhead{$\alpha_{X-ray}$} & \colhead{$\delta_{X-ray}$}} 

\startdata
J0851+3331 & $132.91194$ & $33.51837$ & $132.909^{+0.002}_{-0.0006}$ & $33.521^{+0.001}_{-0.0005}$ \\
J0928+2031 & $142.01891$ & $20.52919$ & $142.0181^{+0.0003}_{-0.0003}$ & $20.5269^{+0.0003}_{-0.0003}$ \\
J0952+3434 & $148.16761$ & $34.57948$ & $148.1679^{+0.0005}_{-0.0005}$ & $34.5771^{+0.0006}_{-0.0006}$ \\
J0957+0509 & $149.41329$ & $5.15884$ & $149.4142^{+0.0006}_{-0.0006}$ & $5.1592^{+0.0004}_{-0.0004}$ \\
J1002+2031 & $150.61184$ & $20.51718$ & $150.6141^{+0.0005}_{-0.0005}$ & $20.5179^{+0.0004}_{-0.0004}$ \\
J1038+4849 & $159.68158$ & $48.82160$ & $159.6792^{+0.0003}_{-0.0003}$ & $48.822^{+0.0002}_{-0.0002}$ \\
J1050+0017 & $162.66625$ & $0.28533$ & $162.6667^{+0.0003}_{-0.0003}$ & $0.2858^{+0.0004}_{-0.0004}$ \\
J1110+6459 & $167.57379$ & $64.99665$ & $167.5742^{+0.0009}_{-0.0009}$ & $64.9975^{+0.0004}_{-0.0004}$ \\
J1138+2754 & $174.53732$ & $27.90854$ & $174.5381^{+0.0003}_{-0.0003}$ & $27.9059^{+0.0004}_{-0.0004}$ \\
J1152+3312 & $178.00078$ & $33.22826$ & $178.003^{+0.001}_{-0.002}$ & $33.2324^{+0.0007}_{-0.0009}$ \\
J1152+0930 & $178.19748$ & $9.50410$ & $178.1982^{+0.0005}_{-0.0005}$ & $9.5039^{+0.0006}_{-0.0006}$ \\
J1207+5254 & $181.89964$ & $52.91645$ & $181.8948^{+0.0008}_{-0.0007}$ & $52.9127^{+0.0007}_{-0.0005}$ \\
J1209+2640 & $182.34864$ & $26.67961$ & $182.3476^{+0.0004}_{-0.0003}$ & $26.6782^{+0.0002}_{-0.0002}$ \\
J1226+2149 & $186.71298$ & $21.83120$ & $186.7131^{+0.0002}_{-0.0002}$ & $21.8314^{+0.0002}_{-0.0002}$ \\
J1226+2152 & $186.71541$ & $21.87372$ & $186.7156^{+0.0002}_{-0.0002}$ & $21.8732^{+0.0002}_{-0.0002}$ \\
J1329+2243 & $202.39392$ & $22.72106$ & $202.3938^{+0.0005}_{-0.0004}$ & $22.7229^{+0.0005}_{-0.0005}$ \\
J1336$-$0331 & $204.00041$ & $-3.52500$ & $204.0013^{+0.0004}_{-0.0004}$ & $-3.5245^{+0.0004}_{-0.0004}$ \\
J1343+4155 & $205.88685$ & $41.91762$ & $205.8869^{+0.0003}_{-0.0003}$ & $41.9176^{+0.0002}_{-0.0002}$ \\
J1420+3955 & $215.16798$ & $39.91924$ & $215.1685^{+0.0004}_{-0.0004}$ & $39.9198^{+0.0003}_{-0.0003}$ \\
J1439+1208 & $219.79075$ & $12.14043$ & $219.7906^{+0.0003}_{-0.0003}$ & $12.1415^{+0.0002}_{-0.0002}$ \\
J1456+5702 & $224.00359$ & $57.03903$ & $224.0025^{+0.001}_{-0.0008}$ & $57.0393^{+0.0005}_{-0.0005}$ \\
J1522+2535 & $230.71986$ & $25.59097$ & $230.7195^{+0.0005}_{-0.0005}$ & $25.5932^{+0.0004}_{-0.0004}$ \\
J1531+3414 & $232.79423$ & $34.24019$ & $232.7942^{+9e-05}_{-0.0001}$ & $34.24008^{+8e-05}_{-9e-05}$ \\
PSZ1G311 & $237.52959$ & $-78.19169$ & $237.5222^{+0.0008}_{-0.0008}$ & $-78.1911^{+0.0002}_{-0.0002}$ \\
J1621+0607 & $245.38488$ & $6.12200$ & $245.3835^{+0.0008}_{-0.0008}$ & $6.123^{+0.001}_{-0.0004}$ \\
J1632+3500 & $248.04523$ & $35.00967$ & $248.044^{+0.002}_{-0.002}$ & $35.010^{+0.001}_{-0.001}$ \\
J1723+3411 & $260.90064$ & $34.19948$ & $260.9008^{+0.0004}_{-0.0004}$ & $34.1992^{+0.0004}_{-0.0004}$ \\
J2243$-$0935 & $340.83636$ & $-9.58859$ & $340.8395^{+0.0005}_{-0.0005}$ & $-9.5954^{+0.0003}_{-0.0002}$ \\
\enddata
\tablecomments{We only report errors in the X-ray centroids since $\alpha$ and $\delta$ are determined to within $\lesssim 1''$ precision}

\end{deluxetable*}


\subsubsection{Concentration Parameter and Centroid Separation}\label{sec:c-d}

The ICM cores of relaxed clusters are expected to be concentrated and aligned with the other observable components. We use the concentration parameter (\emph{c}) to define the level of central core concentration and the separation of the X-ray centroid and BCG in units of kpc (\emph{D} [kpc]) to characterize the alignment.

High central concentrations are associated with the formation of a cool-core \citep{santos2008}. The lack of a cool core is an indication of a disturbed cluster since it suggests that not enough time has passed since recent merger activity for a cool-core to have formed. Although a good indicator of disturbedness when not present, the presence of a cool core may not always indicate that a cluster is relaxed. Some studies argue that mergers typically destroy the cool core \citep[e.g.,][]{allen2001,sanderson2006}, others argue that only more extreme forms of mergers can destroy cool cores \citep[e.g.,][]{markevitch2007,million2010,ehlert2011,mann2012}, and others argue that that cool cores can persist despite merger activity\citep[e.g.,][]{rasia2015,biffi2016,ghirardini2022}. In any case, the concentration parameter has been well tested and found to efficiently differentiate cluster populations especially when combined with a parameter like \emph{A} or \emph{log(w)}. We opt for the definition of concentration used by \cite{cassano2010}, described by equation~\ref{eq:c}.

\begin{equation}
c_{[R_{500}]} = \frac{\mathrm{Flux}(r < 0.2\,R_{500})}{\mathrm{Flux}(r < R_{500})}.
	\label{eq:c}
\end{equation}

In equation \ref{eq:c}, \(\mathrm{Flux}(r < 0.2\,R_{500})\) represents the sum of X-ray counts within the inner \(0.2\,R_{500}\), and \(\mathrm{Flux}(r < R_{500})\) is the total X-ray flux within \(R_{500}\). This concentration parameter provides a simple, non-parametric measure of the central concentration of the X-ray surface brightness profile. A higher concentration typically indicates a denser, more relaxed core. We adopted this form of the concentration parameter since the use of an adaptive physical radius \(R_{500}\) is important for analyzing samples over a wide range of redshifts and allows for more consistent comparisons with other studies \citep[e.g.,][]{hallman2011}.

The connection between the centrality of the BCG with respect to ICM and dynamical state was proposed by \cite{forman1982,jones1984,jones1999}. Alignment between these X-ray emissions and the BCG is expected for relaxed clusters \citep{allen1995,hashimoto2008} but not for all systems \citep{vandenbosch2005,coziol2009,skibba2011,sehgal2013,lauer2014,hoshino2015}. Even though the BCG has been known to wobble \citep{kim2017, harvey2017} and the ICM has been known to slosh \citep{markevitch2001,churazov2003,johnson2010,johnson2012,harvey2017}, the difference in centroid between the ICM distribution and the BCG has still been found to be a robust test of cluster dynamical state since the offsets due to merger activity are much more prominent than residual wobbling/sloshing \citep{mann2012,rossetti2016,zenteno2020}.

It is worth noting that \cite{deluca2021} found that the difference in the X-ray and BCG centroids is more sensitive to dynamical state than the difference between the X-ray peak and the BCG centroid. Though not a direct claim of these papers, this is also supported by \cite{zenteno2020} and \cite{mann2012} which found that centroid separations $D>71$ kpc and peak separations $D>42$ kpc indicated a cluster was undergoing a major merger. The higher threshold associated to the difference in centroids as opposed to the difference in peaks avoids complications where physical effects (i.e wobbling or sloshing) and statistical uncertainties can become more impactful. Thus, we define our \emph{D} [kpc] parameter as described in equation~\ref{eq:D}.

\begin{align}
D\,[\mathrm{kpc}] = d_A \Big[ &\left( (\alpha_{\mathrm{ICM}} - \alpha_{\mathrm{BCG}}) 
\cos(\delta_{\mathrm{BCG}}) \right)^2 \nonumber \\
&+ (\delta_{\mathrm{ICM}} - \delta_{\mathrm{BCG}})^2 \Big]^{1/2} 
\label{eq:D}
\end{align}

Here $\alpha$ and $\delta$ are the right ascension and declination in radians and $d_A$ is the angular diameter distance in units of kpc. The positions of the BCGs were typically easily identified by selecting the unambiguously prominent galaxy cluster member. In cases where there were a number of viable BCG candidates, we used SDSS data to analyze the color-magnitude/color-color space of the cluster field to identify the BCG. The centroid of the ICM was determined from the best fit parameters outputted after modeling the distribution with one or two elliptical Gaussian distributions using the \texttt{CIAO Sherpa} modeling functions. The position coordinates for both the BCG and X-ray centroid are listed in Table~\ref{tab:ra-dec}.

\subsubsection{Parameter Uncertainties}

For \emph{D} [kpc], the errors are derived from the uncertainty in both the X-ray and BCG centroids. Though they are added in quadrature, the uncertainty in the X-ray centroid dominates the total uncertainty. Thus, the errors for \emph{D} [kpc] are effectively determined by the $1\sigma$ deviation around the most likely centroid for the large scale X-ray distribution, found using the cash fitting statistic \citep{Cash1979}. 

For \emph{A}, \emph{log(w)}, and \emph{c} we consider the error due to Poisson noise, the uncertainty in $R_{500}$, and the uncertainty in X-ray centroid. We bootstrap the uncertainties due to each of these factors and add them in quadrature to get a description of the total uncertainty of each of these parameters. For the uncertainty due to Poisson noise, we generate 1000 Poisson noise realizations of our X-ray images and remeasure our parameters. For the uncertainty due to $R_{500}$/the X-ray centroid, we remeasure our parameters 1000 times, randomly selecting $R_{500}$/the X-ray centroid within the bounds of the uncertainty in $R_{500}$/the X-ray centroid. The median percentage error contributions of each source of uncertainty are listed in Table \ref{tab:errs}. We find that the errors in the \emph{c} and \emph{log(w)} parameters are dominantly sourced by the Poisson noise, with \emph{log(w)} having a noticeable sensitivity to the uncertainty of $R_{500}$ and both parameters being only barely affected by the uncertainty in the X-ray centroid. \emph{A} is more sensitive to both the uncertainty in $R_{500}$ and Poisson noise while also showing sensitivity to the X-ray centroid. Overall, \emph{A} is relatively more uncertain/noisy, but remains in this study due to its effectiveness in probing asymmetry in all regions of the cluster.

\begin{deluxetable}{c|ccc}
\setlength{\tabcolsep}{.012\columnwidth}

\tablecaption{The median percentage contribution of each source of uncertainty to \emph{c}, \emph{log(w)}, and \emph{A}. The sources of uncertainty considered are the uncertainty due to the Poisson Noise ($\sigma_{pois}$), the uncertainty in $R_{500}$ ($\sigma_{R_{500}}$), and the uncertainty in the X-ray centroid ($\sigma_{cent}$). \label{tab:errs}}

\tablehead{\colhead{$par$} &  \colhead{$(\sigma_{pois}^2/\sigma_{tot}^2)$} & \colhead{$(\sigma_{R_{500}}^2/\sigma_{tot}^2)$} & \colhead{$(\sigma_{cent}^2/\sigma_{tot}^2)$}}

\startdata
\emph{c} & 81.3$\%$ & 12.4$\%$ & 1.4$\%$ \\
\emph{log(w)} & 56.3$\%$ & 29.2$\%$ & 3.7$\%$ \\
\emph{A} & 32.1$\%$ & 41.0$\%$ & 17.8$\%$ \\
\enddata
\end{deluxetable}


\begin{deluxetable}{ccccc}
\setlength{\tabcolsep}{.012\columnwidth}

\tablecaption{$L_{par}$ is the cut values distinguishing between relaxed and disturbed clusters for each individual parameter. The reference for where $L_{par}$ was derived is listed. The 1st, 2nd (the median, $m_{par}$), and 3rd quartiles of each morphological parameter are listed as calculated from the morphological parameters measured for the whole sample. \label{tab:cuts}}

\tablehead{\colhead{$par$} &  \colhead{$L_{par}$} & \colhead{$q_{1}$} & \colhead{$m_{par}$} & \colhead{$q_{3}$}} 

\startdata
\emph{c} & 0.275 \citep{rasia2013} & 0.16 & 0.20 & 0.27 \\
\emph{log(w)} & -2.229 (this work) & -2.44 & -2.23 & -2.09 \\
\emph{A} & 1.15 \citep{rasia2013} & 1.05 & 1.28 & 1.36 \\
\emph{D} [kpc] & 25 \cite{markevitch2001} & 14.6 & 26.3 & 38.7 \\
\enddata
\tablecomments{For \emph{D} [kpc] we exclude the data point corresponding to SDSS J2243$-$0935}

\end{deluxetable}

\subsection{Combined Parameter}\label{sec:M}

We use a combined parameter, \emph{M}, that incorporates the parameters detailed in sections \ref{sec:a-w} and \ref{sec:c-d} to find the most relaxed and most disturbed clusters \citep{cassano2010,rasia2013,meneghetti2014,lovisari2017,cialone2018,bartalucci2019,deluca2021}. Since the chosen parameters have been shown to perform well in the literature, we opt to define our M parameter by equation~\ref{eq:M} as in \cite{rasia2013} as opposed to giving different parameters different weights based on their efficiency as in \cite{cialone2018}.

\begin{equation}
M=\frac{1}{N_{par}}\sum_{par}{(A_{par}\times\frac{par-L_{par}}{|q_{par}-m_{par}|})}
\label{eq:M}
\end{equation}

\begin{figure*}
	\includegraphics[width=.85\textwidth]{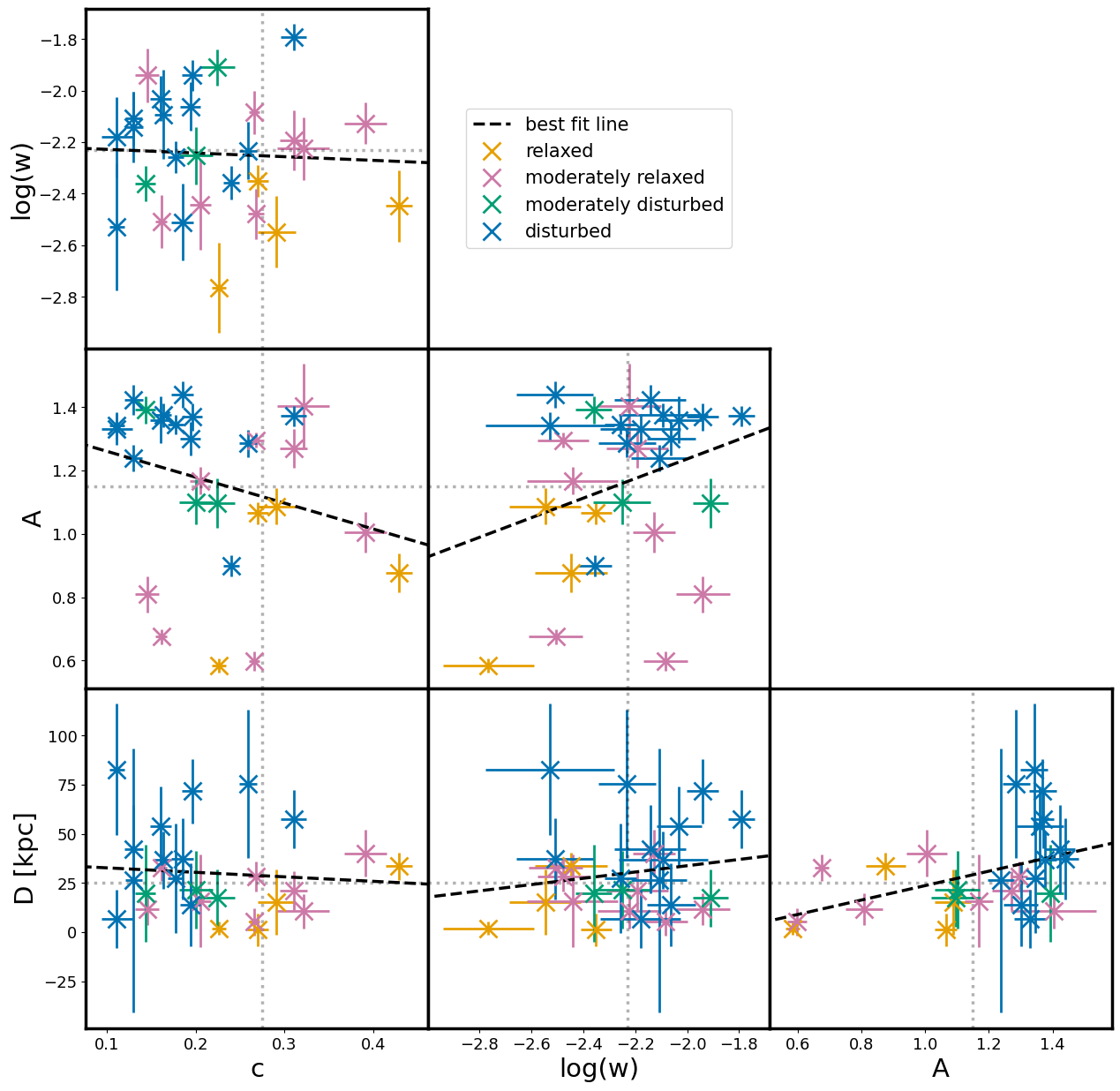}\centering
    \caption{Scatter plots of the various morphological parameters considered in this work. The vertical and horizontal dotted lines indicate the $L_{par}$ cut each individual parameter uses to differentiate between relaxed and disturbed objects. The colors denote the relaxation classification of the cluster with orange = ``relaxed", magenta = ``moderately relaxed", green = ``moderately disturbed", and blue = ``disturbed." We exclude data points associated to SDSS J2243$-$0935 in the panels involving D. The dashed line represent the best fit line.}
    \label{fig:param_comp}
\end{figure*}

In equation~\ref{eq:M}, $par$ is a variable indicating the parameter we are considering, $N_{par}$ is the number of parameters (in our case $N_{par}=4$), $A_{par}$ regulates the sign ($A_{c}=+1$ and $A_{par}=-1$ otherwise, ensuring smaller \emph{D} [kpc],\emph{A}, and \emph{w} values and a larger \emph{c} value results in a larger $M$ value which corresponds to a relaxed cluster), $m_{par}$ is the median, $q_{par}$ is the 1st quartile if $par<m_{par}$ or 3rd quartile value if $par>m_{par}$, and $L_{par}$ is the cut value that differentiates relaxed and disturbed clusters. Table \ref{tab:cuts} lists the values of $L_{par}$. Authors typically either use the median value of the distribution of parameter measurements or some optimized cut to differentiate between relaxed and disturbed clusters. Authors using an optimized cut require a definition of true dynamical state. Generally, they either compare their measurements to simulation results \citep[e.g.,][]{cialone2018} or to a by-eye inspection \citep[e.g.,][]{rasia2013} to find an optimal cut to separate out dynamical state bimodally. As noted by \cite{cao2021}, the $L_{par}$ values are varied and potentially arbitrary; still, an author's chosen cut values are sufficient in identifying the most relaxed/disturbed objects and when combining multiple parameters the intermediate clusters are more accurately classified bimodally. For three of the parameters, we chose values of $L_{par}$ derived from other samples in the literature since we know that our sample is not the representative sample of the underlying cluster population given that it is a strong lensing only sample. The only parameter for which we chose the median value is \emph{log(w)}. We did this since, as noted by \cite{yuan2020}, the \emph{log(w)} parameter can be sensitive to data preprocessing, specifically binning and smoothing. Since we are unaware of a previous representative cluster sample that calculates \emph{log(w)} using the same version of preprocessed images (unsmoothed and binned by $4''\times4''$), we use our median value to distinguish dynamical state. We also note that we do not use the cut derived from \cite{mann2012}, since it was used to find extremely/definitely disturbed objects. We also do not use the cut value associated to the scaled radius ($r=0.02*$$R_{500}$) as in \cite{rossetti2016} and \cite{sanderson2009b}, since these focus studies on the separation of X-ray peak compared to the BCG centroid. Instead, we use the half the value of the average sloshing amplitude postulated by \cite{markevitch2001} which is given by the estimated average length of the cool filament extending from the BCG, $~35 h^{-1}$ kpc. For $H_0 = 70 \mathrm{km}$ $\mathrm{s}^{-1}$ $\mathrm{Mpc}^{-1}$ this yields $L_{D}=25$ kpc. This is close to our median offset value and the maximum sloshing values found in \cite{johnson2010}.


Our results are largely insensitive to our choice in cut values. For $c$, $A$, and $D_y$ [kpc], we vary $L_{par}$ from literature motivated cut-value to the median value of the parameter, both of which are listed in Table \ref{tab:cuts}. For $log(w)$, we vary from $L_{log(w)}$ (derived from the median of our sample) listed in \ref{tab:cuts} to a literature motivated $L_{log(w)}=-2$ \citep{bohringer2010,weismann2013,rasia2013}. In these ranges, the average change in $M$ based on a change in cut ($\overline{\Delta M}$) is less than the average error in the $M$ measurement $\overline{M_{err}}\simeq0.45$. The average changes are $\overline{\Delta M}_c\simeq0.34$, $\overline{\Delta M}_{log(w)}\simeq0.37$, $\overline{\Delta M}_A\simeq0.24$, and $\overline{\Delta M}_{D[kpc]}\simeq0.04$.

Though the changes due to choice of cut are small, they do affect the sign of the $M$ value for 3 of our clusters, thereby changing their dynamical state classification. J1110+6459 ($M=-0.028(0.5)$) flips to a positive $M$ value when varying any of the cut values and J1420+3955 ($M=-0.25(0.3)$) flips to a positive $M$ value when varying any of the cut values other than $L_{D_y[kpc]}$. J1329$+$2243 ($M=-0.41(0.5)$) flips to a positive $M$ value in addition to J1110+6459 and J1420+3955 when varying $L_{log(w)}$. It is within the error bounds for all these objects to transition from relaxed to disturbed and vice versa. However, the transition to $L_{log(w)}=-2$ over-emphasizes $log(w)$, due to our limited sample size a relatively low $log(w)$ value for J1329+2243. $log(w)$ is still a powerful tool for differentiating between relaxed and disturbed objects but due to sensitivity to preprocessing \citep{yuan2020} and our limited $log(w)$ range, the use of the median value is optimal for this study. 

\begin{deluxetable*}{cccccccc}
\centerwidetable

\def\arraystretch{1.2}
\tablecaption{Measurements of the morphological parameters, combined parameter, and final dynamical state classification. The Last column indicates the parameters that identified the cluster as relaxed. \label{tab:measurs}}
\tablehead{
\colhead{Name} & \colhead{\emph{c}} & \colhead{\emph{log(w)}} &  \colhead{A} & \colhead{\emph{D} [kpc]} & \colhead{\emph{M}} & \colhead{classification} & \colhead{relaxed parameters}}

\startdata
J0851+3331 & {$0.26\pm0.01$} & {$-2.2\pm0.1$} & {$1.29\pm0.04$} & {$75\pm38$} & {$-1.3\pm0.8$} & disturbed & log(w)\\
J0928+2031 & {$0.268\pm0.009$} & {$-2.5\pm0.1$} & {$1.30\pm0.02$} & {$28\pm7$} & {$0.20\pm0.24$} & M-relaxed & log(w)\\
J0952+3434 & {$0.13\pm0.01$} & {$-2.1\pm0.1$} & {$1.42\pm0.05$} & {$42\pm22$} & {$-1.3\pm0.5$} & disturbed & \\
J0957+0509 & {$0.14\pm0.01$} & {$-2.36\pm0.07$} & {$1.39\pm0.04$} & {$20\pm25$} & {$-0.42\pm0.5$} & M-disturbed & log(w), D [kpc]\\
J1002+2031 & {$0.163\pm0.009$} & {$-2.1\pm0.2$} & {$1.38\pm0.04$} & {$37\pm15$} & {$-1.1\pm0.4$} & disturbed & \\
J1038+4849 & {$0.161\pm0.006$} & {$-2.5\pm0.1$} & {$0.68\pm0.02$} & {$32\pm6$} & {$1.3\pm0.2$} & M-relaxed & log(w), A\\
J1050+0017 & {$0.29\pm0.02$} & {$-2.5\pm0.1$} & {$1.09\pm0.06$} & {$15\pm17$} & {$1.1\pm0.5$} & relaxed & c, log(w), A, D [kpc]\\
J1110+6459 & {$0.20\pm0.02$} & {$-2.3\pm0.1$} & {$1.10\pm0.07$} & {$22\pm20$} & {$-0.028\pm0.5$} & M-disturbed & log(w), A, D [kpc]\\
J1138+2754 & {$0.31\pm0.01$} & {$-1.79\pm0.05$} & {$1.37\pm0.03$} & {$57\pm15$} & {$-1.2\pm0.3$} & disturbed & c\\
J1152+0930 & {$0.20\pm0.01$} & {$-2.4\pm0.2$} & {$1.17\pm0.04$} & {$16\pm23$} & {$0.29\pm0.6$} & M-relaxed & log(w), D [kpc]\\
J1152+3312 & {$0.111\pm0.008$} & {$-2.5\pm0.2$} & {$1.34\pm0.04$} & {$83\pm33$} & {$-1.5\pm0.8$} & disturbed & log(w)\\
J1207+5254 & {$0.20\pm0.01$} & {$-1.94\pm0.06$} & {$1.37\pm0.04$} & {$72\pm16$} & {$-1.9\pm0.4$} & disturbed & \\
J1209+2640 & {$0.39\pm0.02$} & {$-2.13\pm0.08$} & {$1.00\pm0.06$} & {$40\pm12$} & {$0.68\pm0.4$} & M-relaxed & c, A\\
J1226+2149 & {$0.266\pm0.006$} & {$-2.08\pm0.08$} & {$0.60\pm0.03$} & {$5\pm7$} & {$1.8\pm0.2$} & M-relaxed & A, D [kpc]\\
J1226+2152 & {$0.14\pm0.01$} & {$-1.9\pm0.1$} & {$0.81\pm0.06$} & {$11\pm8$} & {$0.45\pm0.3$} & M-relaxed & A, D [kpc]\\
J1329+2243 & {$0.18\pm0.01$} & {$-2.5\pm0.1$} & {$1.44\pm0.04$} & {$37\pm21$} & {$-0.41\pm0.5$} & disturbed & log(w)\\
J1336$-$0331 & {$0.32\pm0.03$} & {$-2.2\pm0.1$} & {$1.4\pm0.1$} & {$10\pm8$} & {$0.28\pm0.3$} & M-relaxed & c, D [kpc]\\
J1343+4155 & {$0.27\pm0.01$} & {$-2.35\pm0.06$} & {$1.06\pm0.03$} & {$1\pm8$} & {$0.93\pm0.2$} & relaxed & log(w), A, D [kpc]\\
J1420+3955 & {$0.22\pm0.02$} & {$-1.91\pm0.07$} & {$1.10\pm0.08$} & {$17\pm15$} & {$-0.25\pm0.4$} & M-disturbed & A, D [kpc]\\
J1439+1208 & {$0.31\pm0.01$} & {$-2.2\pm0.1$} & {$1.27\pm0.06$} & {$21\pm9$} & {$0.11\pm0.3$} & M-relaxed & c, D [kpc]\\
J1456+5702 & {$0.19\pm0.01$} & {$-2.06\pm0.09$} & {$1.30\pm0.05$} & {$14\pm21$} & {$-0.45\pm0.5$} & disturbed & D [kpc]\\
J1522+2535 & {$0.16\pm0.01$} & {$-2.03\pm0.09$} & {$1.36\pm0.07$} & {$54\pm21$} & {$-1.5\pm0.5$} & disturbed & \\
J1531+3414 & {$0.226\pm0.008$} & {$-2.8\pm0.2$} & {$0.58\pm0.02$} & {$2\pm3$} & {$2.9\pm0.3$} & relaxed & log(w), A, D [kpc]\\
PSZ1G311 & {$0.43\pm0.01$} & {$-2.4\pm0.1$} & {$0.88\pm0.06$} & {$33\pm6$} & {$1.9\pm0.4$} & relaxed & c, log(w), A\\
J1621+0607 & {$0.177\pm0.009$} & {$-2.26\pm0.06$} & {$1.34\pm0.03$} & {$28\pm28$} & {$-0.59\pm0.6$} & disturbed & log(w)\\
J1632+3500 & {$0.130\pm0.009$} & {$-2.1\pm0.1$} & {$1.24\pm0.04$} & {$26\pm67$} & {$-0.81\pm1.4$} & disturbed & \\
J1723+3411 & {$0.11\pm0.02$} & {$-2.2\pm0.2$} & {$1.33\pm0.05$} & {$7\pm15$} & {$-0.51\pm0.4$} & disturbed & D [kpc]\\
J2243$-$0935 & {$0.240\pm0.009$} & {$-2.36\pm0.06$} & {$0.90\pm0.03$} & {$154\pm14$} & {$-1.9\pm0.3$} & disturbed & log(w), A\\
\enddata
\tablecomments{In this table, cluster classifications in the “moderate” categories are abbreviated. For example, “M-relaxed” stands for “moderately relaxed” and “M-disturbed” for “moderately disturbed.”}
\end{deluxetable*}

\section{Results}\label{sec:results}

\subsection{Morphological classification}\label{sec:morph-cls}

In Table~\ref{tab:measurs}, we list the results of our measurements. We additionally list the parameters that identify the cluster in question as relaxed. We note that some parameters Table~\ref{tab:measurs} identify a cluster as relaxed while other parameters may identify the same cluster as disturbed. This suggests that using a singular parameter to characterize the dynamical state of a cluster can be insufficient. Although \emph{M} is more robust against misidentification, it is still not a definitive distinction due to its sensitivity to the chosen $L_{par}$ values, the characteristics of the cluster sample, and the input morphological parameters used to construct it.

To robustly define the dynamical state of our cluster sample, we choose the following classification system to distinguish clusters:
\begin{itemize}
  \item clusters with $M>1$ and at least 3 morphological measurements identifying the cluster as dynamically relaxed are labeled ``relaxed."
  \item clusters with $M>1$ and at least two morphological measurements defining the cluster as relaxed are labeled ``moderately relaxed." The specific combination of \emph{A} and \emph{log(w)} does not count since it can misclassify a cluster in cases of symmetric substructure which is the case for SDSS J2243$-$0935.
  \item clusters with $M<1$ and at least two morphological measurements defining the cluster as relaxed are labeled ``moderately disturbed." The same rule for the specific combination of \emph{A} and \emph{log(w)} also applies here.
  \item clusters not satisfying any of these conditions are labeled ``disturbed."
\end{itemize}

These rules and their associated labels will be used throughout the remainder of this paper. The moderately disturbed/relaxed objects are intermediate clusters that lean either more relaxed or disturbed whereas the relaxed/disturbed clusters have more confident dynamical state determinations. Future scientific work requiring relaxed or disturbed objects should reference these delineations when using clusters from this sample. The specific scientific goals of the future work will determine whether clusters in the two ``moderate" categories should be included in their respective relaxed/disturbed classifications or should be excluded from the work if it requires unambiguously relaxed/disturbed clusters.

\begin{deluxetable}{ccccc}
\setlength{\tabcolsep}{.1\columnwidth}
\tablecaption{Pearson Correlation of the Various Morphological Parameters. \label{tab:corrs}}

\tablehead{\colhead{$par$} &  \colhead{r} & \colhead{p}}

\startdata
\emph{log(w)} vs. \emph{c} & -0.05 & 0.799 \\
\emph{A} vs. \emph{c} & -0.26 & 0.184 \\
\emph{D} [kpc] vs. \emph{c} & -0.09 & 0.672 \\
\emph{A} vs. \emph{log(w)}  & 0.28 & 0.155 \\
\emph{D} [kpc] vs. \emph{log(w)}  & 0.17 & 0.402 \\
\emph{D} [kpc] vs. \emph{A} & 0.43 & 0.027 \\
\enddata
\tablecomments{For parameters involving \emph{D} [kpc] we exclude the data point corresponding to SDSS J2243$-$0935.}
\end{deluxetable}

\subsection{Correlation between Morphological Parameters}\label{sec:morph-corr}

In Figure~\ref{fig:param_comp} we compare the various parameters with each other. Since our sample is not large enough to be statistically robust against general scatter, it is difficult to find strong correlations between the various parameters as denoted in Table~\ref{tab:corrs}. In both Figure~\ref{fig:param_comp} and Table~\ref{tab:corrs}, we remove the measurements associated to SDSS J2243$-$0935 from panels/rows involving $D$ since this is an extremely disturbed object that skews the visualization/analysis of the correlations. 

The strongest correlation is between \emph{A} and \emph{D} [kpc], showing a moderate correlation. The correlation could be in part due to centroid offsets in some clusters, caused by an irregular ICM shape as opposed to a uniform-elliptical object shifted out of alignment with the BCG and other cluster components. \emph{D} [kpc] also shows a very weak correlation with \emph{log(w)} as well and no correlation with \emph{c} suggesting that the centroid offset caused by mergers may have a weak connection to the formation of substructure and may not be strongly related to cool-core destruction. This could be in consequence of the fact that \emph{c} forms in absence of disruptions and has the potential to survive merger activity. Therefore, late-time mergers can occur that result in substructure formation and an offset of the cluster component centroids without destroying the cool core.


\emph{A} shows weak correlations with \emph{log(w)} and \emph{c}. Graphically, we can resolve weak linear correlations despite significant scatter likely due to the noisier nature of the \emph{A} parameter. The \emph{A} vs. \emph{log(w)} and  \emph{A} vs. \emph{c} show a concentration of points around the best fit lines with a large fraction of outliers. Even though \emph{log(w)} and \emph{c} show correlations with \emph{A}, they do not show a correlation with one another. An explanation as to why \emph{A} is weakly related to \emph{log(w)} and \emph{c} even though these parameters are uncorrelated to each other is that \emph{A} is sensitive to asymmetry throughout the cluster, including in the core. \emph{Log(w)} is sensitive to substructure outside the core (beyond $0.15{\times}R_{500}$) which \emph{A} also probes; however, highly concentrated/symmetric cool cores will influence \emph{A}, causing the correlation to \emph{c}. This suggests that \emph{A} is a useful parameter, but may be influenced by both substructure and core density effects simultaneously, leading to additional noise. Since \emph{log(w)} and \emph{c} probe independent spaces, the lack of strong correlations is unsurprising. Again, this highlights that the cool core may be able to survive mergers that result in substructure and X-ray centroid sloshing.

\begin{figure*}
	\includegraphics[width=.85\textwidth]{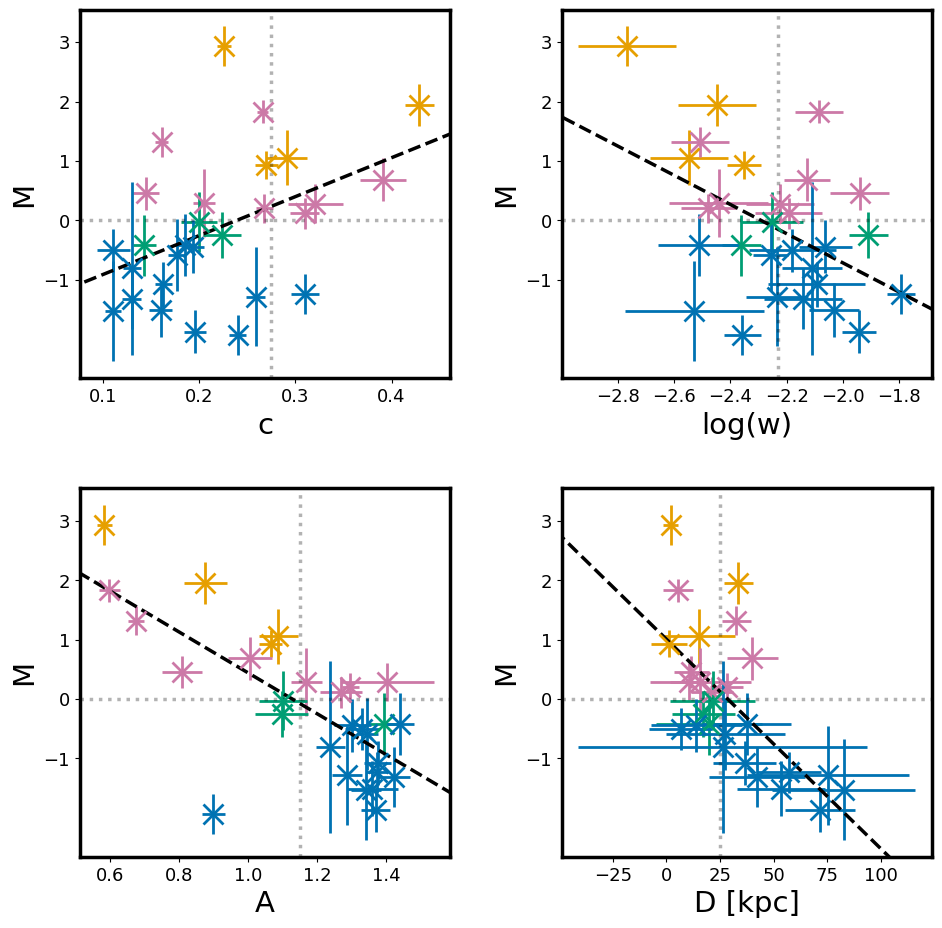}\centering
    \caption{Scatter plots of each of the morphological parameters compared to the combined morphological parameter M. The vertical and horizontal dashed lines indicate the $L_{par}$ cut each individual parameter uses to differentiate between relaxed and disturbed objects. $L_{M} = 0$ by definition. The colors denote the relaxation classification of the cluster with orange =``relaxed", magenta =``moderately relaxed", green =``moderately disturbed", and blue =``disturbed." We exclude data points associated to SDSS J2243$-$0935 in the panels involving D. The dashed line represent the best fit line.}
    \label{fig:M_comp}
\end{figure*}

\subsection{Correlation to the M parameter}\label{sec:M-corr}
In Figure~\ref{fig:M_comp}, we compare the various parameters to the combined parameter \emph{M}. We can see that the parameters have relatively strong correlations to the \emph{M} parameter and thus the final dynamical state as displayed in Table~\ref{tab:corrs_M}.

\begin{figure}
	\includegraphics[width=\columnwidth]{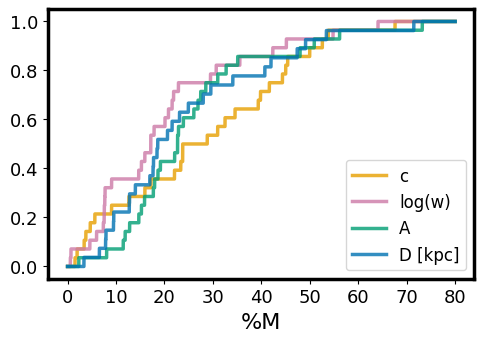}\centering
    \caption{Cumulative Distribution Functions of the percent contribution to the \emph{M} parameter of each of the individual morphological measures.}
    \label{fig:per_cdf}
\end{figure}

It is unsurprising that these morphological parameters would be correlated to the \emph{M} parameter since \emph{M} is built from them. Different parameters will contribute more/less to the final determination of \emph{M} even when we give them equal weights as emphasized in Table~\ref{tab:corrs_M}. This is a consequence of using literature motivated cut-values as opposed to the median values in our sample which would result in more equal contributions based on the construction of $M$. As previously discussed, the literature cut values are well motivated and varying from the median values to the chosen cut values does not significantly change our results. Still, this choice creates a slight preference for cool core clusters being identified as relaxed. On average, the \emph{c} and \emph{A} parameters contribute slightly more than the \emph{log(w)} and \emph{D} [kpc] parameters. We note that \emph{D} [kpc]'s contribution to M slightly declines when omitting SDSS J2243$-$0935. The level of contribution varies drastically from cluster to cluster, although in Figure~\ref{fig:per_cdf} we see that \emph{log(w)} consistently contributes the least, though the difference in its contribution relative to the other parameters is not extreme. Since \emph{log(w)} and \emph{A} both probe substructure, our morphological determination of cluster dynamical state still effectively incorporates and balances substructure in our classification scheme.

Using the combined parameter helps us to mitigate the effects of systematic uncertainties on final dynamical state determination and performs better overall than any singular parameter. Using the Kolmogorov-Smirnov (KS) test, we can better understand how well an individual parameter separates relaxed and disturbed clusters. The test's p-value describes the confidence with which we are able to reject the null-hypothesis that the two samples are drawn from the same sample. In the context of our work, we test how well the individual parameter separates out the relaxed and disturbed clusters as determined by the rules outlined in section~\ref{sec:morph-cls} based on their cut value (see Table~\ref{tab:cuts}). We see \emph{A} and \emph{D} [kpc] perform the best at separating out the sample into two unique populations according to the KS-test with \emph{c} performing less proficiently and \emph{w} performing much less. Figure \ref{fig:M_comp}, shows that we actualize the expected correlations with \emph{M} increasing with \emph{c} and decreasing with \emph{A}, \emph{w}, and \emph{D} [kpc] (see equation \ref{eq:M}).

\begin{deluxetable}{ccccc}
\def\arraystretch{1.3}
\setlength{\tabcolsep}{.05\columnwidth}

\tablecaption{The relations of each individual parameter compared to the combined \emph{M} parameter. r and p are the values associated to the Pearson Correlation, represented graphically in Figure~\ref{fig:M_comp}. The KS-p value is the value associated with the two-sample Kolmogorov-Smirnov test with KS-p$<$0.05 rejecting the null-hypothesis that the two-samples are drawn from the same sample. The \%M column represents the average percent contribution of each of the parameters gives to M as calculated in equation~\ref{eq:M} \label{tab:corrs_M}}
\tablehead{
\colhead{$par$} &  \colhead{r} & \colhead{p} &  \colhead{KS-p} & \colhead{\%M}}

\startdata
\emph{c} & 0.44 & 0.018 & 0.114 & $26^{+28}_{-18}$ \\
\emph{w} & -0.47 & 0.012 & 0.635 & $17^{+20}_{-16}$ \\
\emph{A} & -0.74 & 8e-6 & 6e-4 & $23^{+26}_{-15}$\\
\emph{D} [kpc] & -0.66 & 1e-4 & 1e-4 & $20^{+27}_{-20}$\\
\enddata

\tablecomments{For parameter \emph{D} [kpc], we exclude the data point corresponding to SDSS J2243$-$0935 from the calculation of r,p and KS-p. We keep it in the calculation of \%M which only leads to a 1\% difference.}


\end{deluxetable}

\subsection{Individual Morphological Parameter Correlation with Mass, Radius, and Redshift}\label{sec:comp-morph}

\begin{figure*}
	\includegraphics[width=\textwidth]{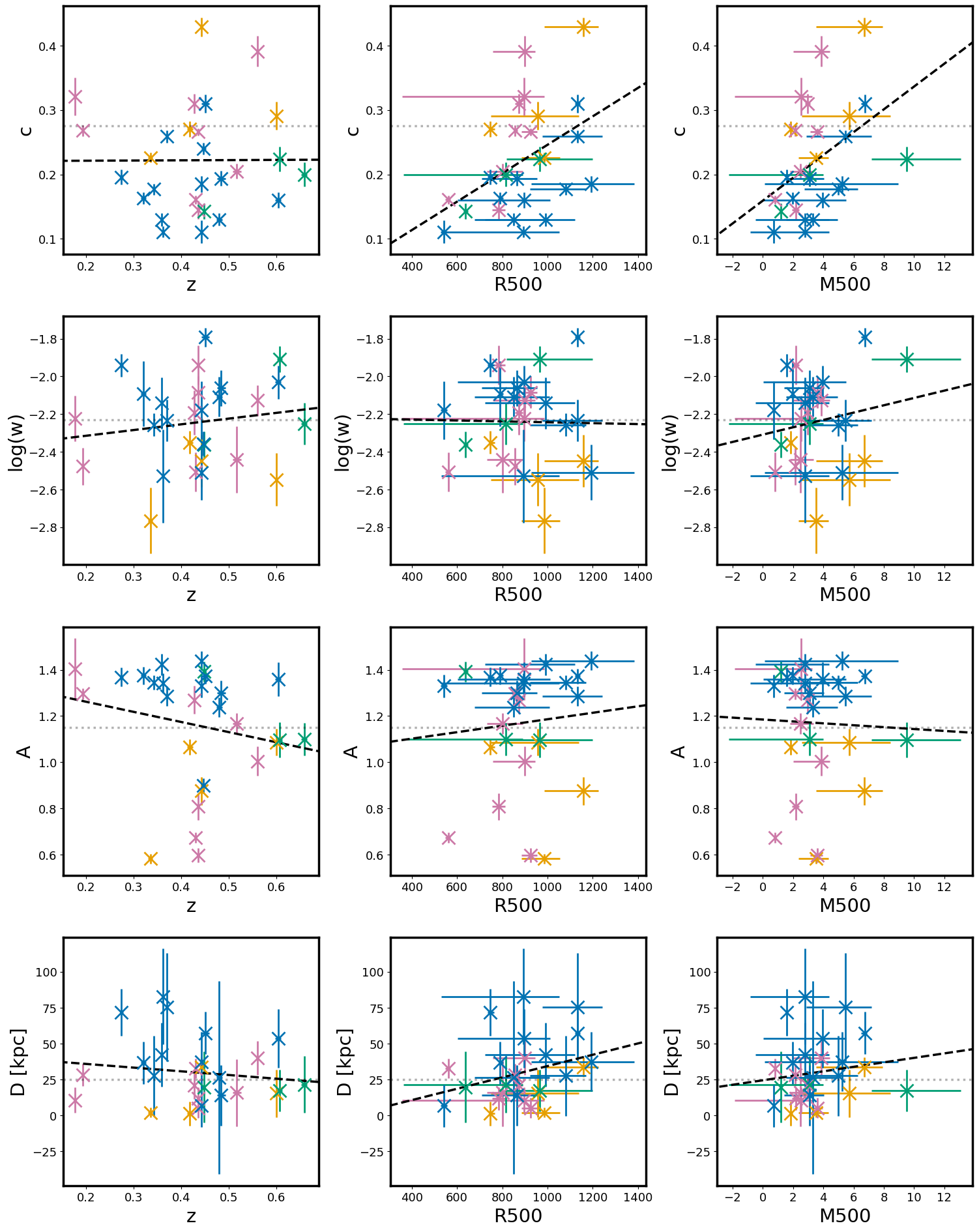}\centering
    \caption{Scatter plots of each morphological parameter vs. redshift, radius, and mass. The colors denote the relaxation classification of the cluster with orange = ``relaxed", magenta = ``moderately relaxed", green = ``moderately disturbed", and blue = ``disturbed." We exclude data points associated to SDSS J2243$-$0935 in the panels involving D, $R_{500}$, and $M_{500}$. The dashed line represent the best fit line.}
    \label{fig:comp_comp}
\end{figure*}

In Table~\ref{tab:comp-corrs} and Figure~\ref{fig:comp_comp} we analyze if the individual morphological measurements correlates with redshift, $R_{500}$, or $M_{500}$ ($M_{500}$ is the total mass enclosed within $R_{500}$). In the following discussion, we removed SDSS J2243$-$0935 from the graphs/calculations associated to D, $R_{500}$, and $M_{500}$ since it is an extreme outlier for all these parameters.

\subsubsection{Redshift}
According to their Pearson correlation coefficients, no parameter shows a very strong correlation to redshift. Graphically, this lack of correlation can easily be confirmed for \emph{c}. \emph{D} [kpc] and \emph{log(w)} have very small r-values and graphically do not see significant signs of correlation. For \emph{A}, we can see a concentration of points around the best fit line, potentially hinting at a weak increase in the value of \emph{A} with a decrease in redshift. 

When we split each value by the median redshift, we reinforce the lack of correlation for \emph{c} and \emph{log(w)}. For \emph{A} and \emph{D} [kpc], the average values between high and low redshift bins differ mildly though with a large degree of scatter. Also, the correlation of these individual parameters imply that clusters become more disturbed in the lower redshift bin. It should be noted that the uncertainties on the binned average of one redshift bin overlaps with the average of the other. In total, the redshift correlation to these parameters is likely insignificant, but cannot be ruled out. 

\subsubsection{Mass and Radius}
To constrain correlations between the parameters and their cluster size we use $R_{500}$. Correspondingly, we use $M_{500}$ to track the correlation with the cluster's mass. These parameters are not independent. As such the qualitative conclusions for trends in mass and radius match for all comparisons in this study. For completeness and comparability to studies that analyze only one of the parameters, we include parameter comparisons to both $M_{500}$ and $R_{500}$. 

From their measured Pearson correlation coefficients, graphically, and by separating them around the median value for $R_{500}$/$M_{500}$, we see that \emph{A} and \emph{log(w)} show no correlation with cluster size/mass. 

\emph{D} [kpc] has a weak to moderate correlation with the size of the cluster, though it is quite scattered. The increase in centroid offset with increased size is also corroborated by the median values of \emph{D} [kpc] for high and low $R_{500}$ clusters, although each of the averages are well within the error bars of each other. The same general trend is seen for $M_{500}$ except the Pearson correlation is much lower. Overall, we cannot rule out the scaling of \emph{D} [kpc] with respect to cluster mass and size.

\emph{c} shows the strongest correlation out of any of the other parameters, although it is still only moderate. The median values of the high and low $R_{500}$/$M_{500}$ clusters reflect the trends seen graphically: the concentration increases as a function of cluster size/mass. The errors of the high and low redshift bin averages do not have as much of an overlap as for other parameters. \cite{bartalucci2019} found a slight (almost negligible) positive correlation between between \emph{c} and $M_{500}$ (r=$0.11\pm0.08$ with a p-value of $0.19\pm0.31$), much less than our r$=0.43$ correlation with p$=0.024$. This may hint at a unique concentration bias for massive strong lensing clusters. The trend for \emph{c} to increase with cluster radius/mass could be a consequence cool cores in strong lensing clusters surviving merger activity; thus persisting in massive clusters that form after multiple mergers.

\begin{deluxetable*}{c|ccccc}
\setlength{\tabcolsep}{20pt}
\def\arraystretch{1.3}

\tablecaption{The correlation of each individual parameter and the combined \emph{M} parameter to the cluster properties: redshift, radius, and mass. r and p are the values associated to the Pearson Correlation represented graphically in figures~\ref{fig:comp_comp} and~\ref{fig:comp_comp_M}. We also calculate the average value for each morphological parameter for a sub sample of clusters either above or below the median value of the given cluster property. 
\label{tab:comp-corrs}}
\tablehead{
\colhead{$prop$} & \colhead{$par$} & \colhead{r} & \colhead{p} & \colhead{$m_{par}$} & \colhead{$m_{par}$}\\
\colhead{} & \colhead{} & \colhead{} & \colhead{} & \colhead{[$prop<m_{prop}$]} & \colhead{[$prop>m_{prop}$]} }

\startdata
{} & \emph{c} & 4e-3 & 0.982 & $0.21^{+0.07}_{-0.06}$ & $0.20^{+0.1}_{-0.05}$ \\
{} & \emph{log(w)} & 0.15 & 0.439 & $-2.2^{+0.2}_{-0.3}$ & $-2.2^{+0.2}_{-0.2}$\\
{$z$} & \emph{A} & -0.20 & 0.318 & $1.3^{+0.08}_{-0.4}$ & $1.2^{+0.2}_{-0.2}$ \\
{} & \emph{D} [kpc] & -0.14 & 0.498 & $27^{+30}_{-17}$ & $22^{+19}_{-9}$ \\
{} & \emph{M} & 0.05 & 0.786 & $0.16^{+1.3}_{-1.3}$ & $-0.42^{+1.1}_{-0.82}$\\
\hline
{} & \emph{c} & 0.44 & 0.021 & $0.18^{+0.07}_{-0.04}$ & $0.26^{+0.09}_{-0.08}$ \\
{} & \emph{log(w)} & -0.02 & 0.933 & $-2.2^{+0.2}_{-0.2}$ & $-2.2^{+0.2}_{-0.3}$ \\
{$R_{500}$} & \emph{A} & 0.09 & 0.655 & $1.3^{+0.07}_{-0.2}$ & $1.3^{+0.09}_{-0.4}$ \\
{} & \emph{D} [kpc] & 0.30 & 0.134 & $21^{+22}_{-12}$ & $34^{+22}_{-20}$ \\
{} & \emph{M} & -4e-3 & 0.986 & $-0.22^{+0.8}_{-0.9}$ & $-0.25^{+1.7}_{-0.8}$\\
\hline
{} & \emph{c} & 0.43 & 0.024 & $0.18^{+0.08}_{-0.04}$ & $0.23^{+0.1}_{-0.06}$ \\
{} & \emph{log(w)} & 0.17 & 0.4 & $-2.2^{+0.2}_{-0.2}$ & $-2.2^{+0.2}_{-0.3}$ 
\\
{$M_{500}$} & \emph{A} & -0.03 & 0.873 & $1.3^{+0.07}_{-0.2}$ & $1.1^{+0.2}_{-0.3}$ \\
{} & \emph{D} [kpc] & 0.14 & 0.478 & $20^{+25}_{-11}$ & $28^{+24}_{-15}$ \\
{} & \emph{M} & 0.03 & 0.863 & $-0.15^{+0.83}_{-1.0}$ & $0.25^{+1.7}_{-0.8}$\\
\enddata
\tablecomments{We exclude data points associated to SDSS J2243$-$0935 in the panels involving D, $R_{500}$, and $M_{500}$.}
\end{deluxetable*}

\begin{figure*}
	\includegraphics[width=\textwidth]{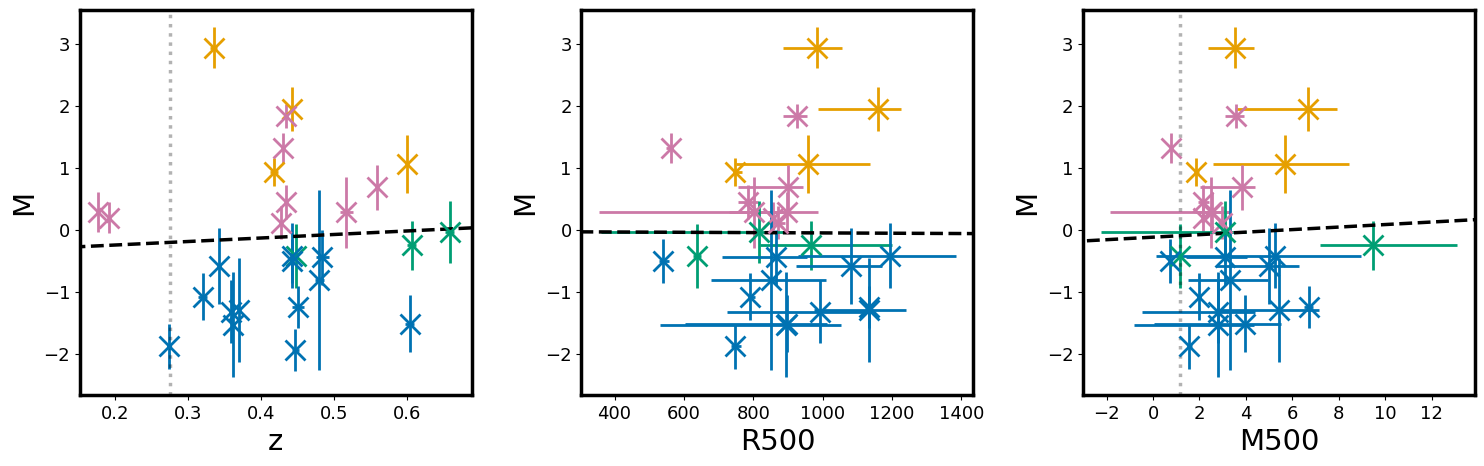}\centering
    \caption{The combined \emph{M} parameter vs. redshift, radius, and mass. The colors denote the relaxation classification of the cluster with orange = ``relaxed", magenta = ``moderately relaxed", green = ``moderately disturbed", and blue = ``disturbed." We exclude data points associated to SDSS J2243$-$0935 in the panels involving $R_{500}$ and $M_{500}$. The dashed line represent the best fit line.}
    \label{fig:comp_comp_M}
\end{figure*}

\section{Discussion}\label{sec:disc}
\subsection{Dynamical Relaxation Correlation with Mass, Radius, and Redshift}\label{sec:comp-M}

In section~\ref{sec:comp-morph}, we compared the individual parameters to investigate potential correlations with respect to mass, radius, and redshift. To fully test dynamical state dependency on these cluster properties, we should use the combined parameter \emph{M} since it is a more representative measurement of the cluster's true underlying dynamical state. The results of this comparison are shown in Figure~\ref{fig:comp_comp_M} and Table~\ref{tab:comp-corrs}.

\begin{figure*}
	\includegraphics[width=\textwidth]{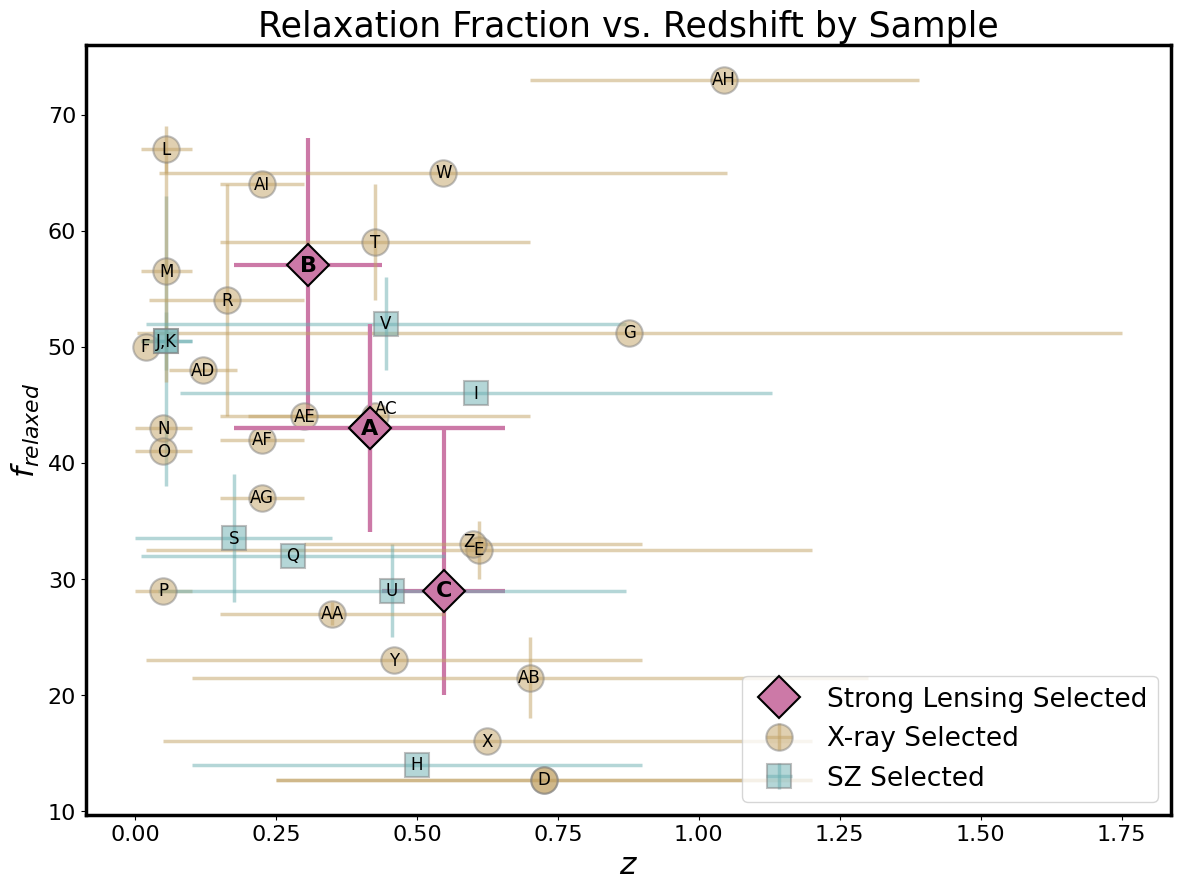}\centering
    \caption{Relaxation fraction measurements compiled from various works, including the cluster samples detailed in Table 5 of \citealt{deluca2021}, and additional data from \citealt{ghirardini2022}, \citealt{campitiello2022}, and this work. The marker labels in the plot correspond to the ``Label" column in Table \ref{tab:rel_frac}. The plotted redshift values represent the central redshift of each sample, computed as the midpoint: ($z=(z_{min}+z_{max})/2$). The error-bars on the x-axis represent the redshift range of the corresponding study. Additional details about the redshift distributions can be found in the original studies listed in the ``Paper” column of Table \ref{tab:rel_frac}.}
    \label{fig:rel_frac}
\end{figure*}

\subsubsection{Redshift}\label{sec:z-M}
In the literature, there is no clear consensus on whether the frequency of relaxed/disturbed clusters changes with redshift. Early studies found that the fraction of relaxed clusters increases at lower redshifts \citep[e.g.,][]{jeltema2005,andersson2009}. Later studies did not find a significant correlation \citep[e.g.,][]{mantz2015,nurgaliev2017,mcdonald2017,ghirardini2022}.

In our work, we do not find a definite correlation between the redshift and the total dynamical state characterized by \emph{M}. We analyze \emph{M}'s correlation with redshift graphically and also by splitting the sample in half using the median redshift ($z=0.4385$). When comparing the high and low $z$ sample, we find a median value of $m_{M}=0.16^{+1.35}_{-1.3}$ for low $z$ and $m_{M}=-0.42^{+1.05}_{-0.82}$ for high z. Though the median values for the high $z$ are lower on average, the scatter of values is quite large. Additionally, the selection of high vs. low redshift is sample specific and as arbitrary as any other cut. Graphically, we can see that shifting the cut value would significantly change the average \emph{M} value for high and low samples. Overall, these measurements cannot rule out/confirm correlations due to their statistical limitations.

The hierarchical formation model suggests that clusters should have a higher fraction of relaxed clusters at lower redshifts \citep{bohringer2010}. \cite{mcdonald2017} found a lack of correlation in their relaxation fraction and postulated that it could be due to the relaxation time of the hot gas being proportional to the crossing time.

\cite{bartalucci2019} found a very weak decrease with their combined \emph{M} parameter. Though our sample does not show a significant correlation between dynamical state and redshift, our binned sample suggests that dynamical state could change with redshift, although caveats due to binning selection, our limited redshift range, and our limited sample size still remain. In total, larger samples satisfactorily probing the complete mass and redshift space are needed to track whether relaxation changes with redshift.

\begin{deluxetable*}{cccccc}
\centerwidetable
\setlength{\tabcolsep}{8pt}
\def\arraystretch{1.2}
\tablecaption{Summary of relaxation fraction measurements for various cluster samples with their own selection method, redshift range, and sample size. The asterisk symbol (*) next to the citation in the Paper column indicates that the relaxation fraction was calculated in \cite{deluca2021} instead of directly presented in the original study. The letter labels in the first column match the letter labels on the markers in Figure~\ref{fig:rel_frac}.\label{tab:rel_frac}}
\tablehead{
\colhead{Label} & \colhead{Paper} & \colhead{Number of Objects} & \colhead{$z_{min}$} & \colhead{$z_{max}$} & \colhead{Relaxation Fraction}
}
\startdata
A  &  This work (Full sample)  &  28  &  0.176  &  0.656  &  $43^{+9}_{-9}$ \\
B  &  This work (Low z)  &  14  &  0.176  &  0.4385  &  $57^{+11}_{-13}$ \\
C  &  This work (High z)  &  14  &  0.4385  &  0.656  &  $29^{+14}_{-9}$ \\
D  &  \cite{campitiello2022}$^a$  &  118  &  0.25  &  1.2  &  $13$ \\
E  &  \cite{ghirardini2022} (High z)  &  325  &  0.02  &  1.2  &  $33^{+3}_{-3}$ \\
F  &  \cite{ghirardini2022} (Low z)  &  325  &  0.017  &  0.02  &  $50$ \\
G  &  \cite{yuan2020}  &  964  &  0.003  &  1.75  &  $51$ \\
H  &  \cite{zenteno2020}  &  288  &  0.1  &  0.9  &  $14$ \\
I  &  \cite{bartalucci2019}*  &  74  &  0.08  &  1.13  &  $46$ \\
J  &  \cite{lopes2018} (SZ sample, X-ray Indicators)  &  40  &  0.01  &  0.1  &  $51^{+3}_{-3}$ \\
K  &  \cite{lopes2018} (SZ sample, Optical Indicators)  &  40  &  0.01  &  0.1  &  $51^{+13}_{-13}$ \\
L  &  \cite{lopes2018} (X-ray sample, X-ray Indicators)  &  62  &  0.01  &  0.1  &  $67^{+2}_{-2}$ \\
M  &  \cite{lopes2018} (X-ray sample, Optical Indicators)  &  62  &  0.01  &  0.1  &  $57^{+10}_{-10}$ \\
N  &  \cite{chon2017} (Flux Limited Sample 2)  &  42  &  \nodata  & \ensuremath{<0.1}  &  $43$ \\
O  &  \cite{chon2017} (Flux Limited Sample 1)  &  51  &  \nodata  &  \ensuremath{<0.1}  &  $41$ \\
P  &  \cite{chon2017} (Volume Limited Sample)  &  93  &  \nodata  &  \ensuremath{<0.1}  &  $29$ \\
Q  &  \cite{lovisari2017}*  &  120  &  0.01  &  0.55  &  $32$ \\
R  &  \cite{andradesantos2017} (X-ray Sample)  &  100  &  0.025  &  0.3  &  $54^{+10}_{-10}$ \\
S  &  \cite{andradesantos2017} (SZ Sample)  &  164  &  \nodata  &  \ensuremath{<0.35}  &  $34^{+6}_{-6}$ \\
T  &  \cite{rossetti2017} (X-ray Sample)  &  104  &  0.15  &  0.7  &  $59^{+5}_{-5}$ \\
U  &  \cite{rossetti2017} (SZ Sample)  &  169  &  0.04  &  0.87  &  $29^{+4}_{-4}$ \\
V  &  \cite{rossetti2016} (SZ Sample)  &  132  &  0.02  &  0.87  &  $52^{+4}_{-4}$ \\
W  &  \cite{lavoie2016}  &  85  &  0.043  &  1.05  &  $65$ \\
X  &  \cite{mantz2015}  &  361  &  0.05  &  1.2  &  $16$ \\
Y  &  \cite{parekh2015}  &  84  &  0.02  &  0.9  &  $23$ \\
Z  &  \cite{nurgaliev2013}  &  36  &  0.3  &  0.9  &  $33$ \\
AA  &  \cite{mahdavi2013}*  &  50  &  0.15  &  0.55  &  $27$ \\
AB  &  \cite{maughan2012}*  &  114  &  0.1  &  1.3  &  $22^{+4}_{-4}$ \\
AC  &  \cite{mann2012}*  &  108  &  0.15  &  0.7  &  $44$ \\
AD  &  \cite{bohringer2010}*  &  31  &  0.06  &  0.18  &  $48$ \\
AE  &  \cite{cassano2010}*  &  32  &  0.2  &  0.4  &  $44$ \\
AF  &  \cite{zhang2010}*/\cite{okabe2010}*  &  12  &  0.15  &  0.3  &  $42$ \\
AG  &  \cite{sanderson2009b}  &  65  &  0.15  &  0.3  &  $37$ \\
AH  &  \cite{santos2008} (High z)  &  15  &  0.7  &  1.39  &  $73$ \\
AI  &  \cite{santos2008} (Low z)  &  11  &  0.15  &  0.3  &  $64$ \\
\enddata
\tablecomments{a: Many clusters in \cite{campitiello2022} were not assigned a dynamical state (62 mixed clusters). As such, the quoted relaxation fraction is a lower limit using their ``very relaxed" dynamical state label.}
\end{deluxetable*}

\subsubsection{Mass and Radius}\label{sec:$M_{500}$-M}

The correlation between dynamical state with mass has also shown inconclusive results. Simulation work predicts that more massive clusters are more disturbed on average \citep[e.g.,][]{gouin2021,kuchner2020}. In the hierarchical formation model, more massive structures require more interaction to form; thus, more massive clusters at a given redshift should be more disturbed then their low mass counterparts \citep{giocoli2012}. Other studies find that cluster dynamical state has no dependence on cluster mass \citep[e.g.,][]{nurgaliev2017,bartalucci2019}.

Graphically, we do not see any significant correlation with mass when considering the entire sample. However, we see a mild correlation when comparing the high mass and low mass subsamples split according to the median mass ($M_{500}=3.09*10^{14} \mathrm{M_{\odot}}$). We find the median \emph{M} values are $m_{M}=-0.15^{+0.83}_{-1.0}$ for low $M_{500}$ and $m_{M}=0.25^{+1.7}_{-0.8}$ for high $M_{500}$. The overlap in uncertainty is still very large and the mass range is limited. To more accurately test whether total mass is correlated to dynamical state, a study would need a sufficiently large sample of clusters at approximately the same redshift.

The radius of the cluster, $R_{500}$, is intimately related to the mass of the cluster, $M_{500}$, since both are derived from the same density condition. We do not see a change in relaxedness due to changing $R_{500}$ both graphically and when comparing the high and low radius subsamples split at the median radius ($R_{500}=893.2$ kpc). We find the median \emph{M} values are $m_{M}=-0.22^{+0.83}_{-0.86}$ for low $R_{500}$ and $m_{M}=-0.25^{+1.78}_{-0.8}$ for high $R_{500}$. 

Though the dynamical state does not correlate with radius, the choice of aperture does affect most parameters as shown in the literature (e.g., \cite{deluca2021}). Different areas of the cluster are more sensitive to dynamical activity, usually with the outskirts being more dynamically active. Therefore, a fixed aperture would give inconsistent measurements for clusters of different sizes; hence, a scaled aperture like $R_{500}$ should be used.

\subsection{Relaxation Fraction}\label{sec:rel-frac}

Using the relaxation classifications outlined in section~\ref{sec:morph-cls}, we calculate the relaxation fraction, $f_{\mathrm{relaxed}}$, as the ratio of relaxed systems to the total number of clusters in the sample:
\begin{equation}
f_{\mathrm{relaxed}} = 100\%\times\frac{N_{\mathrm{relaxed}}}{N_{\mathrm{total}}}.
\label{eq:f}
\end{equation}
Here, $N_{\mathrm{relaxed}}$ includes clusters classified as either ``relaxed'' or ``moderately relaxed''. The uncertainties in $f_{\mathrm{relaxed}}$ are calculated from the Bayesian binomial confidence intervals drawn from the $1\sigma$ confidence interval for the beta distribution \citep{Cameron2011}.

We first consider the full sample across the redshift range $z=0.176$--$0.656$, and find a relaxation fraction of $f_{\mathrm{relaxed}} = 43\%^{+9}_{-9}$. We also examine changes with redshift by splitting the sample at the median redshift ($z=0.4385$), as in section~\ref{sec:z-M}. For the low-redshift subsample ($z<0.4385$), we find a relaxation fraction of $57\%^{+11}_{-13}$. For the high-redshift subsample ($z>0.4385$), we find $29\%^{+14}_{-9}$. These results suggest a slight increase in the relaxation fraction with redshift, though interpretation is limited by the small sample size and classification uncertainties as discussed in section~\ref{sec:z-M}.

As detailed in \cite{cao2021} and \cite{deluca2021}, the comparison of relaxation fraction between samples is difficult due to the variation in cluster selection functions, mass ranges, redshift ranges, choice in the definition of relaxedness, etc. In Figure \ref{fig:rel_frac} and Table \ref{tab:rel_frac}, we combine information from a variety of studies recorded and tabulated in \citealt{deluca2021} with additional measurements from the CHEX-MATE sample \citep[see][]{campitiello2022}, the eFEDS sample \citep[see][]{ghirardini2022}, and the SGAS sample (this work).


Ignoring the intricacies of each individual sample selection and definition of relaxedness, we get that the average value of relaxation across the papers listed in \cite{deluca2021} + \cite{campitiello2022} and \cite{ghirardini2022} is $\sim44\%$. 
Although this estimate incorporates relaxation fractions derived from clusters with differing selection functions, mass and redshift ranges, and definitions of relaxedness, it is consistent with the estimated relaxation fraction for our sample.

To our knowledge, there is no former study whose selection function is very similar to our sample's selection. \cite{mann2012} ($z=0.15-0.7$) and \cite{mahdavi2013} ($z=0.15-0.55$) have redshift ranges most similar to ours. \cite{mahdavi2013} uses the Canadian Cluster Comparison Project (CCCP) sample. They note this sample's selection function is not entirely understood (based off spatial, temperature, and redshift restrictions), yet is statistically indistinguishable from other clusters samples. They use centroid shift, power ratios, central entropy, and X-ray-optical offset to classify cluster dynamical state. \cite{mann2012} use the most X-ray luminous clusters in the MAssive Cluster Survey (MACS) sample and classify their relaxation by the X-ray peak-bcg and X-ray centroid-bcg separations. \cite{mahdavi2013} found the relaxation fraction to be $27\%$ and \cite{mann2012} the relaxation fraction to be $44\%$. \cite{rossetti2017} also used the MACS sub-sample described in \cite{mann2012} to find a relaxation fraction of $59\pm5\%$ using the concentration parameter. \cite{rossetti2017} uses the same parameter on PSZ1-cosmo \citep{plank2014} with follow up X-ray data to compare to the X-ray selected sample. They find the SZ selected sample is significantly more disturbed with a relaxation fraction of  $29\pm4\%$. This trend of a significant bias towards higher relaxation fractions for X-ray selected samples is found by many authors \citep[e.g.,][]{andradesantos2017,chon2017,lopes2018}. 

Generally, in Figure \ref{fig:rel_frac} and Table \ref{tab:rel_frac}, we see SZ selected samples have lower relaxation fractions that X-ray selected ones. For instance, \citep{andradesantos2017} finds a relaxation fraction of $54\%^{+10}_{-10}$ for their X-ray sample and $34\%^{+6}_{-6}$ for their SZ sample which places our measured $43\%^{+9}_{-9}$ very centrally between these two values for their X-ray and SZ samples. Some X-ray studies yield similar relaxation fractions to the value measured for our subsample (e.g., \cite{mann2012,cassano2010} with relaxation fractions $\sim44\%$), whereas the ROSAT sample show relaxation fractions from X-ray selected samples to be much higher \citep{santos2008}. Some SZ studies show relaxation fractions comparable to our measurement (e.g., the \emph{Planck}+SPT sample considered in \cite{bartalucci2019} yielding a relaxation fraction of $\sim46\%$), although typically purely SZ selected samples yield lower relaxation fractions (e.g., the 288 SPT clusters considered in \cite{zenteno2020}).

As discussed in~\ref{sec:slgc}, a strong lensing selected sample is subject to unique biases. The overall effect those biases have on the average relaxation is largely unexplored. Comparing to studies with similar redshift ranges, we find that our sample is not as disturbed as an SZ selected cluster population but is slightly more disturbed than an X-ray selected sample. Many studies favor SZ as a more representative sample of the underlying cluster population \citep[e.g.,][]{andradesantos2017,chon2017,lopes2018}. Additionally, simulation results from \cite{Cui2017} yield a relaxation fraction of $\sim35\%$. The results from SZ and simulations suggest that the concentration bias may outweigh the merger bias found in strong lensing galaxy clusters, making them on average more relaxed than the underlying cluster population. The prevalence of the concentration bias in strong lensing clusters could help explain the excess of strong lensing features in cluster observations compared to simulations \citep{Meneghetti2020}. The effect is not strong enough to produce a higher relaxation fraction than that observed in X-ray selected samples. The bias towards disturbedness is higher if we ignore the ``moderately relaxed" cluster category.


\begin{figure}
	\includegraphics[width=.95\columnwidth]{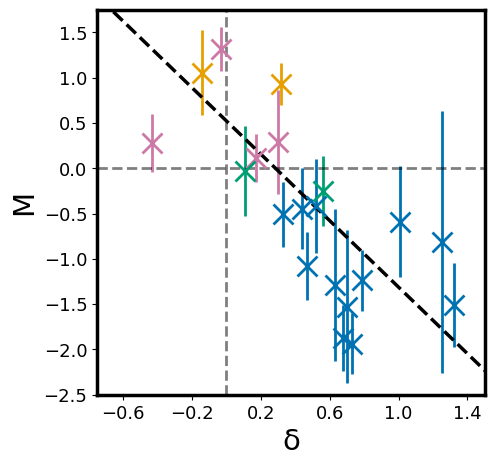}\centering
    \caption{Scatter plot of our combined morphological \emph{M} to the delta parameter \emph{$\delta$} from \cite{yuan2020}.The colors denote the relaxation classification of the cluster with orange = ``relaxed", magenta = ``moderately relaxed", green = ``moderately disturbed", and blue = ``disturbed." The dashed line represent the best fit line.}
    \label{fig:yuan}
\end{figure}

\subsection{Direct Cluster Comparison}\label{sec:yuan}

\cite{yuan2020} calculated the relaxation parameters of 964 clusters using archival data from the Chandra X-ray Observatory. In this large sample, 20 of the 28 clusters in this paper have been assigned a dynamical state. Their custom relaxation parameter, \emph{$\delta$}, determines dynamical state using the cluster's X-ray brightness distribution and its X-ray asymmetry. Positive \emph{$\delta$} values mean the cluster is disturbed. In Figure~\ref{fig:yuan}, we compare our measurements to those published in \cite{yuan2020}. We find a correlation of $r=-0.65$ with associated p-value of $2$e-4 between their \emph{$\delta$} parameter and our \emph{M} parameter. We find that the relaxation fraction of this subsample as determined by our methods is $30^{+10}_{-20}$\% whereas \cite{yuan2020} identifies the relaxation fraction of this sub sample to be only $15$\%. Overall, we find fairly good agreement with the \emph{$\delta$} parameter using very different dynamical state determination methods.

\section{Summary}\label{sec:sum}

This paper investigates the dynamical state of 28 strong lensing galaxy clusters at redshifts $z=0.176-0.656$. The goals of this paper were to (A) create a strong lensing galaxy cluster sample with well defined dynamical state measurements; (B) constrain correlations, biases, and disagreements between different morphological measurement proxies and the combined parameter \emph{M} to each other and other cluster properties; (C) calculate and compare our cluster relaxation fraction to other samples. The associated resolution of these goals are as follows:

\renewcommand{\labelenumi}{(A\arabic{enumi})}
\begin{enumerate}
    \item Using the rules outlined in section~\ref{sec:morph-cls}, we find 4 clusters that are definitely relaxed and 13 clusters that are definitely disturbed.
    \item We find 11 clusters that lie in a moderate dynamical state with 8 of those 11 being more relaxed and the other 3 being more disturbed.
    \item We find that our relaxation determination shows a large degree of correlation and overall good agreement with the subsample of clusters measured by \cite{yuan2020}.
\end{enumerate}

\renewcommand{\labelenumi}{(B\arabic{enumi})}
\begin{enumerate}
    \item Most of the parameters are not significantly dependent on one another. There are some mild correlations between \emph{A} and all other parameters, most significantly with \emph{D} [kpc]. \emph{D} [kpc] also may have a weak correlation with \emph{log(w)}.
    \item We do not find any very significant correlation with redshift, though hints of a mild correlation are found for \emph{A}, \emph{D} [kpc], and \emph{M}. Further studies with better statistics are needed to constrain, confirm, or refute this possible correlation.
    \item \emph{c} shows the most significant correlation with mass and radius. A possible explanation is that large structures take longer times to form and strong lensing clusters may be biased cases where major merger activities do not disturb the cool core formation. \emph{D} [kpc] also shows some relation to cluster mass and size, though not as significant/constrained.
\end{enumerate}

\renewcommand{\labelenumi}{(C\arabic{enumi})}
\begin{enumerate}
    \item We find a cluster relaxation fraction of $43\%^{+9}_{-9}$ across our whole sample with a relaxation fraction of $57\%^{+11}_{-13}$ for the low redshift bin and $29\%^{+14}_{-9}$ for the high redshift bin. 
    \item Our relaxation fraction is approximately the average relaxation fraction of $44^{+14}_{-17}$\% across relevant former studies tabulated in \cite{deluca2021} with the addition of \cite{campitiello2022} and \cite{ghirardini2022}.
    \item Our results indicate that a strong lensing selected cluster sample (clusters with larger strong lensing cross-sections at a given mass) have comparable dynamical relaxation fractions to those of mass-selected samples. Our sample is slightly more relaxed than SZ selected cluster samples but not as relaxed as X-ray selected samples.

\end{enumerate}

\section*{Acknowledgements}

R.G. would like to thank the listed coauthors for their support during this project and helpful input when writing the manuscript. This research has made use of data obtained from the Chandra Data Archive provided by the Chandra X-ray Center (CXC). The data set constituting the Chandra Strong Lens Sample,  obtained by the Chandra X-ray Observatory, is hosted at the Chandra Data Collection (CDC) 434~\dataset[doi:10.25574/cdc.434]{https://doi.org/10.25574/cdc.434}. This data set compiles observations associated to Chandra programs GO-19800436 (PI: Bayliss),GO-19700449 (PI: Bayliss), GO-16800783 (PI: Baum), GO-04800884 (PI: Buote), GO-11800671 (PI: Irwin), GO-17800085 (PI: Johnson), GO-12800164 (PI: Morris), GO-17800681 (PI: Rossetti),GO-11800852 (PI: Rykoff), and GTO-03800995 (PI: Vanspeybroeck). Support for this work was provided by the National Aeronautics and Space Administration through Chandra Award Number GO8-19111A issued by the Chandra X-ray Center, operated by the Smithsonian Astrophysical Observatory for and on behalf of the National Aeronautics Space Administration under contract NAS8-03060.

\facilities{CXO (ACIS-I, ACIS-S), HST (WFC3), Sloan.}

\software{This research made use of pandas \citep{McKinney_2010, McKinney_2011}; Sherpa \citep{2001SPIE.4477...76F}; Astropy, a community-developed core Python package for Astronomy \citep{2018AJ....156..123A, 2013A&A...558A..33A}; SciPy \citep{Virtanen_2020}; matplotlib, a Python library for publication quality graphics \citep{Hunter:2007}; ds9, a tool for data visualization supported by the Chandra X-ray Science Center (CXC) and the High Energy Astrophysics Science Archive Center (HEASARC) with support from the JWST Mission office at the Space Telescope Science Institute for 3D visualization; NumPy \citep{harris2020array}; the CIAO X-ray Analysis Software \citep{2006SPIE.6270E..1VF}; Astropy, a community-developed core Python package for Astronomy \citep{2018AJ....156..123A, 2013A&A...558A..33A}; Photutils \citep{Bradley2022}; and pyregion, developed by Jae-Joon Lee (\url{https://github.com/leejjoon}).
}

\bibliography{test}{}
\bibliographystyle{aasjournal}

\end{document}